\documentclass[preprint2]{aastex62}

\def\fer{{{\it Fermi }}\/}

\def\mnras{MNRAS}
\def\apj{ApJ}
\def\aap{A\&A}

\def\apjl{ApJL}
\def\apjs{ApJS}

\shorttitle{Identifying the 3FHL catalog: CTIO results}
\shortauthors{Desai et al.}

\usepackage{natbib}
\usepackage{float}
\usepackage{color}
\usepackage{graphicx}
\usepackage{gensymb}
\usepackage{array}
\usepackage{enumitem}
\usepackage{longtable}
\usepackage{hyperref}

\submitjournal{\apj}
\accepted{01/24/2019}

\begin{document}

\title{Identifying the 3FHL catalog: III. Results of the CTIO-COSMOS optical spectroscopy campaign}

\author{A.Desai}
\affiliation{Department of Physics and Astronomy, Clemson University, Clemson, SC 29634, USA}
\author{S. Marchesi}
\affiliation{Department of Physics and Astronomy, Clemson University, Clemson, SC 29634, USA}
\author{M. Rajagopal}
\affiliation{Department of Physics and Astronomy, Clemson University, Clemson, SC 29634, USA}
\author{M. Ajello} 
\affiliation{Department of Physics and Astronomy, Clemson University, Clemson, SC 29634, USA}

\begin{abstract}
Active galactic Nuclei (AGNs) with their relativistic jets pointed towards the observer, form a subclass of luminous gamma-ray sources commonly known as blazars. The study of blazars is essential to improve our understanding on the AGNs emission mechanisms and evolution, as well as to map the extragalactic background light. To do so, however, one  needs to correctly classify and measure a redshift for a large sample of these sources. The Third {\it Fermi}--LAT Catalog of High-Energy Sources (3FHL) contains $\approx1160$ blazars reported at energies greater than $10$\,GeV. However $\sim$25\% of these sources are unclassified and $\sim$50\% lack of redshift information. To increase the spectral completeness of the 3FHL catalog, we are working on an optical spectroscopic follow up campaign using 4--m and 8--m telescopes. In this paper, we present the results of the second part of this campaign, where we observed 23 blazars using the 4\,$m$ telescope at CTIO in Chile. We report all the 23 sources to be classified as BL Lacs, a confirmed redshift measurement for 3 sources, a redshift lower limit for 2 sources and a tentative redshift measurement for 3 sources.
\end{abstract}


\section{Introduction}

Blazars are a peculiar class of active galactic nuclei (AGNs) which dominate the observable $\gamma$-ray Universe because of their extreme properties and abundant population. The blazar properties are a result of non-thermal emitting plasma traveling towards the observer causing relativistic amplification of flux. This leads to an amplification of low energy photons in the medium to intense levels via inverse Compton process, making blazars valuable sources to understand the physics of an AGN. The Third \fer--LAT Catalog of High-Energy Sources \citep[3FHL][]{ajello17}, which encompasses seven years of observations made by the Large area telescope (LAT) aboard the {\it Fermi Gamma-ray Space Telescope} \citep{atwood09}, contains more than 1500 sources detected at $>10$\,GeV, the vast majority of which ($\approx$ 1160) are blazars \citep{ajello17}.

Innovative scientific results can be obtained using the blazar data collected by the \fer LAT in the $\gamma$-ray regime, provided the redshift ($z$) of the observed blazar source is known. These are not only limited to blazar physics such as, understanding their basic emission processes \citep[e.g. ][]{ghisellini17} or their evolution with redshift \citep{ajello14}, but also to other areas of study, like understanding the extragalactic background light (EBL), which encompasses all the radiation emitted by stars and galaxies and reprocessed radiation from interstellar dust, and its evolution with $z$ \citep{ackermann12,dominguez13}. Out of the confirmed blazar sources reported in the 3FHL catalog a redshift measurement of only $\approx$50\,\% sources is present \citep{ajello17}. To overcome this limitation, extensive optical spectroscopic campaigns, targeting those 3FHL objects still lacking redshift and classification, must be performed.
%are undertaken which make use of observations from ground based telescopes of individual blazar sources.}

Besides being used for redshift determination, optical spectroscopy campaigns of blazars are also essential to distinguish between blazar sub-classes, namely BL Lacs (BLL) and flat spectrum radio quasars (FSRQs). FSRQs are generally high redshift objects with average luminosity larger than that of the BLL \citep{padovani92,paiano17}. As a result, the emission lines in the BLL spectra are weak or absent and the lines in FSRQs are extremely prominent. This is seen by the difference in the equivalent width (EW) of the lines where generally, FSRQ have lines with EW$>5$\r{A} and BLL have lines with  EW$<5$ \r{A}\citep{urry95,ghisellini17}. The blazar sources not classified as FSRQ or BLL are listed as blazar candidates of uncertain type (BCU) in the 3FHL catalog, and constitute $\approx 25\,\%$ of the reported blazar sample \citep{ajello17}. Obtaining a spectroscopically complete classification of the blazars observed by \fer LAT in the $\gamma$-ray regime is essential to validate claims of different cosmological evolution of the two classes \citep{ajello12,ajello14}. 

The ground based telescopes used in the spectroscopy campaigns are generally of the 4--m,8--m and 10--m class type. While the 10--m and 8--m class telescopes are shown to be significantly more effective in obtaining redshift measurements for blazars \citep[60--80\% versus 25--40\% success rate, see, e.g.][]{paiano17,marchesi18}, even 4--m class telescopes have proven to be useful for effectively distinguishing between the two different blazar subclasses \citep[see ][]{shaw13,massaro14,paggi14,landoni15,ricci15,marchesini16,alvarez16a,alvarez16b}.  

This work is part of a larger spectroscopic follow-up campaign to classify the BCUs in the 3FHL catalog and measure their redshift. 
{The first part of the campaign took place in the second half of 2017, when we observed 28 sources in seven nights of observations at the 4--m telescope at Kitt Peak National Observatory (KPNO). The results of this work are reported in \citet{marchesi18}: we classified 27 out of 28 sources as BL Lacs, while the remaining object was found to be a FSRQ. Furthermore, we measured a redshift for 3 sources and set a lower limit on $z$ for other four objects; the farthest object in our KPNO sample has $z>$0.836.
The spectroscopic campaign will then continue with seven nights of observations at the 4--m telescope at Cerro Tololo Interamerican Observatory (CTIO) in Chile and five nights of observations at the 8--m Gemini-N and Gemini-S telescopes (to be performed in  2019). In this work, we report the results} of the observations made during the first four nights at CTIO. Our source sample contains 23 BCUs in the 3FHL catalog without a redshift measurement. The paper is organized as follows: Section~\ref{sample_sel} reports the criteria used in sample selection, Section~\ref{obs} describes the methodology used for the source observation and spectral extraction procedures, Section~\ref{spectral} lists the results of this work, both, for each individual source and also in general terms, while Section~\ref{conclusion} reports the conclusions inferred from this spectroscopic campaign.

\section{Sample Selection}\label{sample_sel}
We selected the 23 objects in our sample among the BCUs in the 3FHL catalog, using the following three criteria.
\begin{itemize}
\item {\bf The object should have an measured optical magnitude measurement}, and it should be V$\le$19.5. Based on previous works, sources with magnitude V$>$19.5 require more than two hours of observations to obtain an acceptable signal-to-noise ratio (S/N), therefore significantly reducing the number of sources that one can observe in a night.
\item The 3FHL source should be bright in the hard $\gamma$-ray spectral regime ($f_{\rm 50-150 GeV}>$10$^{-12}$ erg s$^{-1}$ cm$^{-2}$). Selecting 3FHL objects bright in the 50--150\,GeV band ensures that the completeness of the 3FHL catalog evolves to lower fluxes as more optical observations are performed.
\item The target should  be observable from Cerro Tololo with an altitude above the horizon $\delta$$>$40\,\degree (i.e., with airmass $<$1.5): this corresponds to a declination range -80\degree$<$Dec$<$20\degree. The target should also be observable in October, when the observations take place (i.e., it should have R.A.$\geq$09h0m00s and R.A.$\leq$0h30m00s).
\end{itemize}

\begin{figure*}
  \centering
  \includegraphics[width=0.9\textwidth]{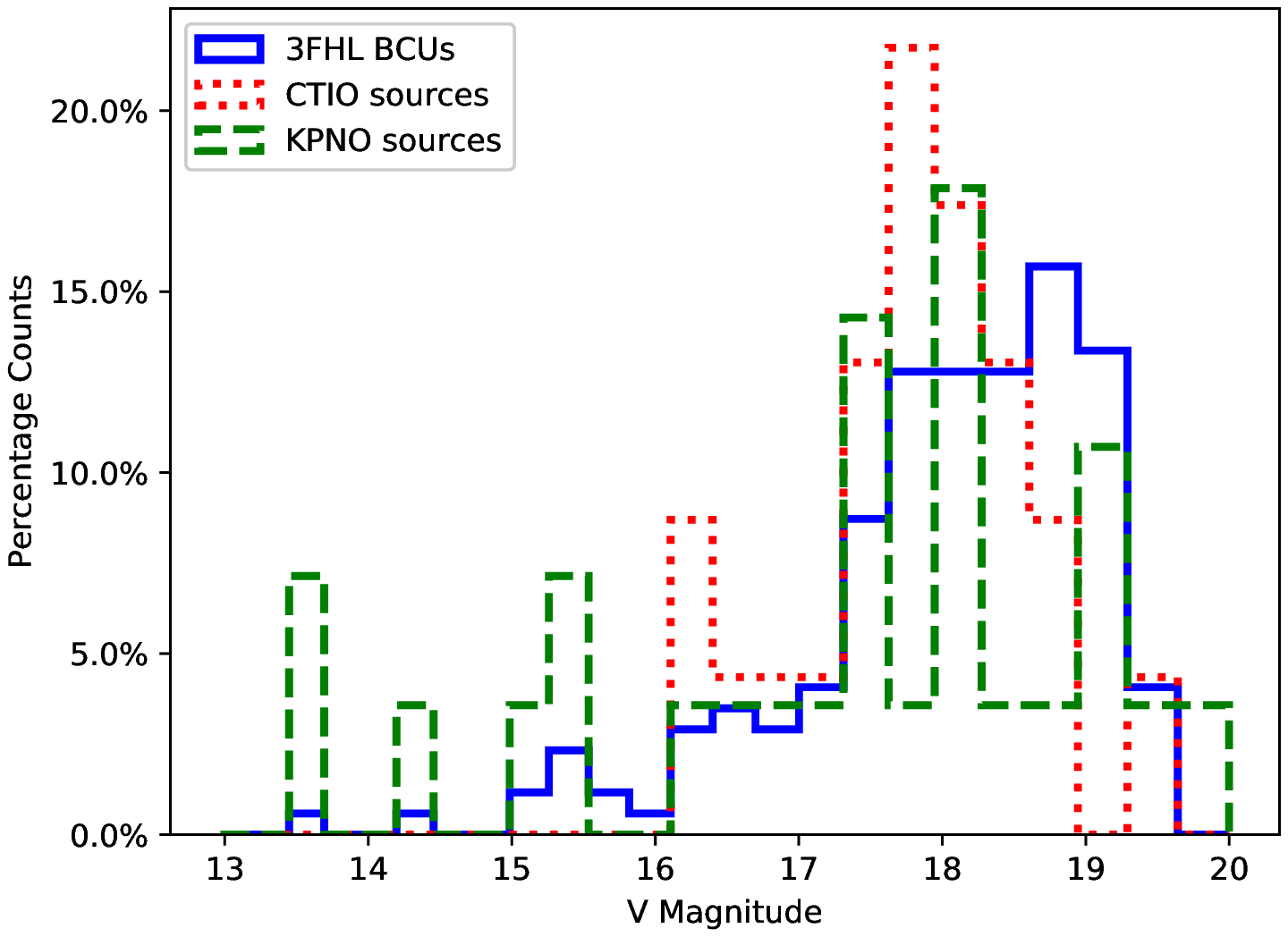}
\caption{{Normalized V-band magnitude distribution of the sources analyzed in this work (red dashed line), compared with the distribution of the 173 3FHL BCUs lacking of redshift and having magnitude information (blue solid line). The magnitude distribution of the objects analyzed in \citet{marchesi18} using KPNO is also shown for comparison.}}
\label{fig:hist}
\end{figure*}

A total of 77 3FHL sources satisfy all these criteria. Our 23 sources were selected among these 77 objects with the goal of covering a wide range of optical magnitudes (V=[16--19.5]) and, consequently, of potential redshifts and luminosities. {In Figure \ref{fig:hist} we show the normalized V-band magnitude distribution of our sources, compared with the one of the overall population of 173 3FHL BCUs still lacking a redshift measurement and having available magnitude information. We also plot the magnitude distribution of the 28 sources studied in \citet{marchesi18}, where we sampled a larger number of bright sources (V$<$16) which all turned out to be featureless BL Lacs.}
The sources used in our sample and their properties are listed in Table~\ref{tab:sample}.

\section{Observations and Data Analysis}
\label{obs}
All the sources in our sample were observed using the 4\,$m$ Blanco telescope located at the Cerro Tololo Inter-American Observatory (CTIO) in Chile. The spectra were obtained using the COSMOS spectrograph with the Red grism and the 0.9$^{\prime\prime}$ slit. This experimental setup corresponds to a dispersion of $\sim$ 4\,\AA\ pixel$^{-1}$, over a wavelength range $\lambda$=[5000--8000]\,\AA, and a spectral resolution R$\sim$2100. The data were taken with the slit aligned along the parallactic angle. %corresponding to $\approx$4\,\AA per pixel dispersion.%This experimental setup corresponds to a dispersion of $\sim$ 4\,\AA\ pixel$^{-1}$.
The seeing was  1.3$^{\prime\prime}$ during the first and third night, 1$^{\prime\prime}$ during the second night and 2.2$^{\prime\prime}$ in the last night, respectively; all four nights were photometric.

All spectra reported here are obtained by combining at least three individual observations of the source with varying exposure times. This allows us to reduce both instrumental effects and cosmic ray contribution. The data reduction is done following a standard procedure: the final spectra are all bias-subtracted, flat-normalized and corrected for bad pixels. { We normalize the flat-field to remove any wavelength dependent variations that could be present in the flat-field source but not in the observed spectrum. This is done by fitting a cubic spline function on the calibration spectrum and taking a ratio of the flat-field to the derived fit \citep[see response function in ][]{iraf_doc}. We choose an order $>$5 for the cubic spline function fit with a $\chi^2$ value less than 1 to account for all variable features in the flat-field }An additional visual inspection is also done on the combined spectra to remove any artificial features that may still be present. This data reduction and spectral extraction is done using the IRAF pipeline \citep{tody86}.

The wavelength calibration for each source is done using the Hg-Ne lamp: we took a lamp spectrum after each observation of a source, to avoid potential shifts in the pixel-$\lambda$ calibration due to changes in the telescope position during the night. Finally, all spectra were flux-calibrated using a spectroscopic standard, which were observed using the same 0.9$^{\prime\prime}$ slit used in the rest of the analysis, %The standard spectrum is obtained two times for each observing night, one in the beginning and other at the end of the night. Once the calibrations are completed, 
and then corrected for the Galactic reddening using the extinction law by \citet{cardelli89} and the $E(B-V)$ value based on the \cite{schlafly11} measurements, as reported in the NASA/IPAC Infrared Science Archive.\footnote{\url{http://irsa.ipac.caltech.edu/applications/DUST/}}

\section{Spectral Analysis}
\label{spectral}

{ To visually enhance the spectral features of our sources, in Figure \ref{fig:spec} we report the normalized spectra of the objects in our sample. These normalized spectra are obtained by dividing the flux-calibrated spectra using a continuum fit \citep[an approach similar to the one reported in ][]{landoni18}.
The continuum is taken to be a power-law unless the optical shape is more complex, in which case the preferred fit is described in ~\ref{individual_src}. The S/N of the normalized spectrum is then measured in a minimum of five individual featureless regions in the spectrum with a width of $\Delta\lambda\approx40$\,\r{A}.} The spectral analysis results for each source, including the computed S/N, are reported in Table~\ref{tab:redshift}.

To find a redshift measurement, each spectrum was visually inspected for any absorption or emission feature. Any potential feature that matched known atmospheric lines\footnote{\url{https://www2.keck.hawaii.edu/inst/common/makeewww/Atmosphere/atmabs.txt}} was not taken into consideration. To test the reliability of any potential feature, its existence was verified in each of the individual spectral files used to obtain the final combined spectrum shown in Fig~\ref{fig:spec}. { For example, the broad emission feature seen in the spectrum of 3FHL J0935.2-1735 around 5633\,\r{A} is not found in the individual files and is thus considered to be an artifact.} The verified features are then matched with common blazar lines,  such as the Mg II doublet lines (2797\,\r{A} and 2803\,\r{A}) or O III line (5007\,\r{A}), to compute the redshift.

All the sources in our sample were classified as BLL based on their spectral properties. Out of the 23 sources, we were able to determine a redshift measurement for 3 sources, a lower limit on the redshift for 2 of them and a tentative redshift measurement for 3 of them. The remaining 15 sources in our sample were found to be featureless. { Details for some of the sources for which a spectral feature or redshift is found are given in Sec~\ref{individual_src}. These features are also listed in Table~\ref{tab:redshift} with the derived redshift measurement.}

\subsection{Comments on Individual sources}
\label{individual_src}

%{\bf3FHL J0002.1-6728:} This BCU is associated with the radio source SUMSS J000215-672653. We were not able to detect any emission or absorption lines in the optical spectrum. It is therefore classified as a featureless BLL.

%{\bf3FHL J0935.2-1735:} This source is associated with the radio source NVSS J093514-173658. {\bf A possible broad emission feature is seen at 5633\r{A}, however it is not detected in all three observed individual spectra. It is, therefore, classified as not real.} \textcolor{blue}{If it's true I'd remove this artifact from the final spectrum, and avoid mentioning the feature here.} No other significant emission or absorption features are seen in the source spectrum. The source is thus classified as a featureless BL Lac without known redshift.

{\bf3FHL J0936.4-2109:} This BCU is associated with the X-ray source 1RXS J093622.9-211031. The optical spectrum of this source shows the presence of two absorption features at 6176\,\r{A} and 6160\,\r{A}. If they are associated with the Mg II doublet, a redshift measurement of 1.1974 and 1.1976 is obtained respectively. Corresponding to this $z$ value, other typical features observed in blazars, either in emission or in absorption (e.g., the Ca II doublet, the G-band, O II or O III features) will fall out of our observed wavelength range of $5000$\r{A}$-8200$\r{A}. We report a tentative lower limit of the redshift as $z>1.197$ for this BLL.

{\bf3FHL J1030.6-2029:} This source is associated with the radio source NVSS J103040-203032. Its optical spectrum shows the presence of the Mg II doublet at 5579\,\r{A} and 5591\,\r{A} respectively. This gives a redshift lower limit of $z>0.995$.

{\bf3FHL J1042.8+0055:} This source is associated with the X-ray source  RBS 0895. A redshift value of 0.73 exists in the literature, \citep{mnras90}, however the authors flagged it as an uncertain measurement. We were not able to detect any absorption or emission lines in our optical spectrum, so we classify this source as a BLL.

%{\bf3FHL J1130.5-7801:} This BCU is associated with the radio source SUMSS J113032-780105. The source is found to be a featureless BLL due to absence of any clear emission or absorption lines.

{\bf3FHL J1155.5-3418:}  This source is associated with the radio source NVSS J115520-341718. The Mg II doublet is identified in the optical spectrum of the source at 5174\,\r{A} and 5185\,\r{A} allowing us to measure the lower limit of the redshift as $z>0.849$.
%and identify the source as a BLL.

{\bf3FHL J1212.1-2328:} This source is associated with the radio source PMN J1212-2327. We obtain an optical spectrum with S/N of 102.8 and detect an emission feature at 8345\,\r{A} with an equivalent width of 0.8\,\r{A}. If associated to the O III line, we derive a redshift $z$=0.666.
%, with the source being classified as a BLL.

{\bf3FHL J1223.5-3033:} This source is associated with the radio source NVSS J122337-303246. We see possible absorption features at 5245\,\r{A}, 5256\,\r{A}, 5577\,\r{A} and 6341\,\r{A}. If 5245\,\r{A} and 5256\,\r{A} absorption features are associated with the Mg II line, a redshift of $0.875$ is measured. However we were not able to detect the presence of any other features and also identify the features at 5577\,\r{A} and 6341\,\r{A} to confirm the redshift measurement with certainty. This source is thus classified as a BLL and a tentative lower limit of $z$$>$0.875 is reported.

%{\bf3FHL J1229.7-5304:} This source is associated with the radio source AT20G J122939-530332. As no potential emission or absorption lines are found in its optical spectrum, this source is classified as a featureless BLL.

%{\bf3FHL J1315.9-0732:} This source is associated with the IR source WISE J131552.98-073301.9. No emission or absorption features are detected in the optical spectrum, allowing us to classify this source as a featureless BLL.

{\bf3FHL J1433.5-7304:} This source is associated with the X-ray source 1RXS J143343.2-730433. One emission feature (H$_\alpha$) and four absorption features (G-band, Mg I,Na and Ca+Fe ) are detected in the spectrum. This gives us a redshift measurement of $z = 0.200$.
%and classify this source as a BLL.

{\bf3FHL J1439.4-2524:} This source is associated with the radio source NVSS J143934-252458. We detect two strong absorption lines at 6008\,\r{A} and 6115\,\r{A} and an absorption line at 6835\,\r{A} close to an atmospheric feature (6845\,\r{A}) in its optical spectrum. If these lines are associated with the Mg I, Ca+Fe and NaD absorption features respectively, a redshift of $z=0.16$ is derived.
%, with the source being classified as a BLL. 

{\bf3FHL J1605.0-1140:} The IR counterpart of this source is WISE J160517.53-113926.8. The optical spectrum shows the presence of an emission feature at 6801\,\r{A} with equivalent width of 7.044\,\r{A}. This feature can be associated with the O II or O III line giving a redshift of 0.824 or 0.358 respectively, however due to no significant detection of any other emission or absorption features and a low S/N measurement, the redshift of this source cannot be measured with certainty. %This source is classified as a BLL.

\section{Conclusion}

{ In this work, we present the results the optical spectroscopic campaign directed towards  rendering the 3FHL a spectroscopically complete sample using the COSMOS spectrograph mounted on the 4\,$m$ Blanco telescope at CTIO in Chile.} We observed 23 extragalactic sources classified as BCU (blazars of uncertain classification) in the 3FHL catalog.

All the objects in our source sample are classified as BLL based on their observed optical spectrum. In the 3FHL catalog, out of the already classified 901 blazars $\approx 84.1\%$ sources are classified as BLL. Moreover out of the 28 sources observed by \cite{marchesi18}, 27 are identified as BLL denoting that our results are not surprising.

Out of the 23 BLL in our sample we find a reliable { redshift} measurement for 3 sources, a reliable { redshift} constraint for 2 sources, a tentative { redshift} constraint for 3 sources and a featureless spectrum with no { redshift} measurement for the remaining 15 sources. Combining our results with the results of \cite{marchesi18}, our optical spectroscopic campaign reports a redshift measurement for $\approx23.5\%$ of the observed BLL sources using 4\,$m$ telescopes. This measurement is in line with the expected consistency of $10-35\%$, obtained for redshift determination of pure BLL using using 4\,$m$ telescopes \citep{landoni15,ricci15,alvarez16a,pena17}. Moreover, our work combined with \cite{marchesi18} also classifies, as either BLL or FSRQs, 51 blazars of previously uncertain classification. 

The third and fourth part of our spectroscopic campaign will include observations from the 4\,$m$ CTIO telescope and 8\,$m$ Gemini-N and Gemini-S telescope respectively\footnote{{\it Fermi} Guest Investigator Program Cycle 11, ID:111128, PI: S. Marchesi.}. Additionally we also aim to extend the campaign by inducing follow up observations\footnote{{\it Swift} Cycle 14, prop ID 1417063 PI: M. Ajello}, similar to \cite{kaur18}, using the Swift X-ray telescope. These follow up observations in the X-ray regime will help us confirm the classification of the blazar sources contributing to the spectral completion of the 3FHL catalog.

\label{conclusion}

\section{Acknowledgements}

A.D. acknowledges funding support from NSF through grant AST-1715256. S.M. acknowledges support from NASA contract 80NSSC17K0503. The authors thank Alberto Alvarez and Sean Points for the help provided during the observing nights at CTIO. This work made use of the TOPCAT software (Taylor 2005) for the analysis of data tables.

\begingroup
\renewcommand*{\arraystretch}{1.8}
\begin{table*}
\centering
\scalebox{0.65}{
\begin{tabular}{lcccccccc}
\hline
\hline
3FHL Name & Counterpart & R.A. & Dec & E(B-V) & mag & Obs Date & Exposure & continuum slope\\
(1) & (2) & (3) & (4) & (5) & (6) & (7) & (8) & (9)\\
\hline
3FHL J0002.1$-$6728 & SUMSS J000215$-$672653 & 00:02:15.21 & $-$67:26:52.91 & 0.0253 & 18.6 & June 1 2018 & 5400 & $-1.44$\\ 
3FHL J0935.2$-$1735 & NVSS J093514$-$173658 & 09:35:14.77 & $-$17:36:58.30 & 0.0643 & 17.8 & June 1 2018 & 3900 & $-0.12$ \\
3FHL J0936.4$-$2109 & 1RXS J093622.9$-$211031 & 09:36:23.08 & $-$21:10:39.00 & 0.0574 & 18.5 & June 2,3 2018 & 5100 & $-0.28$ \\
3FHL J1030.6$-$2029 & NVSS J103040$-$203032 & 10:30:40.46 & $-$20:30:32.70 & 0.0469 & 18.4 &June 3 2018 & 3300 & $-1.91$\\
3FHL J1042.8$+$0055 & RBS 895 & 10:43:03.84 & $+$00:54:20.43 & 0.0419 & 19.3 & June 4 2018 & 5600 & -1.03\\
3FHL J1130.5$-$7801 & SUMSS J113032$-$780105 & 11:30:32.92 & $-$78:01:05.20 & 0.1921 & 17.6 & June 2 2018 & 3400 & $-1.07$\\
3FHL J1155.5$-$3418 & NVSS J115520$-$341718 & 11:55:20.43 & $-$34:17:18.30 & 0.0702 & 16.8 & June 1 2018 & 2400 & $-1.10$\\ 
3FHL J1212.1$-$2328 & PMN J1212$-$2327 & 12:12:04.54 & $-$23:27:42.00 & 0.0656 & 18.2 & June 1 2018 & 4500 & $-0.77$\\
3FHL J1223.5$-$3033 & NVSS J122337$-$303246 & 12:23:37.32 & $-$30:32:46.10 & 0.0593 & 17.2 & June 2 2018 & 3400 & $-2.15$\\ 
3FHL J1229.7$-$5304 & AT20G J122939$-$530332 & 12:29:39.93 & $-$53:03:32.20 & 0.1293 & 17.8 & June 3 2018 & 2300 & $-0.44$\\
3FHL J1315.9$-$0732 & WISE J131552.98$-$073301.9 & 13:15:53.00 & $-$07:33:02.07 & 0.0352 & 18.2 & June 4 2018 & 4500 & $-0.87$\\
3FHL J1433.5$-$7304 & GALEX J143343.0$-$730437   & 14:33:42.81 & $-$73:04:36.84 & 0.1592 & 17.9 & June 1 2018 & 4000 & $-0.81$\\
3FHL J1439.4$-$2524 & NVSS J143934$-$252458 & 14:39:34.66 & $-$25:24:59.10 & 0.0862 & 16.2 & June 3 2018 & 2800 & $-0.01$\\
3FHL J1605.0$-$1140 & WISE J160517.53$-$113926.8 & 16:05:17.53 & $-$11:39:26.83 & 0.2584 & 18.7 & June 4 2018 & 5400 & $-0.35$\\ 
3FHL J1612.3$-$3100 & NVSS J161219$-$305937 & 16:12:19.95 & $-$30:59:37.80 & 0.2003 & 18.1 & June 2 2018 & 3600 & $-1.11$\\
3FHL J1640.1$+$0629 & NVSS J164011$+$062827 & 16:40:11.06 & $+$06:28:27.70 & 0.0695 & 18.6 & June 2 2018 & 3800 & $-1.71$\\
3FHL J1842.4$-$5841 & 1RXSJ184230.6$-$584202   & 18:42:29.67 & $-$58:41:57.19 & 0.0848 & 17.5 & June 1 2018 & 3600 & $-1.67$\\
3FHL J1924.2$-$1548 & NVSS J192411$-$154902 & 19:24:11.82 & $-$15:49:02.10 & 0.1491 & 17.7 & June 3 2018 & 3600 & $-1.35$\\ 
%3FHL J1944.4$-$4523 & 1RXSJ194422.6$-$452326   & 296.0813 & $-$45.3627 & & 15.6 & June 1 2018 & 1080 \\
3FHL J2034.9$-$4200 & SUMSS J203451$-$420024 & 20:34:51.06 & $-$42:00:37.60 & 0.0360 & 17.2 & June 2,4 2018 & 3900 & $-0.62$\\
3FHL J2041.7$-$7319 & SUMSS J204201$-$731911 & 20:42:01.85 & $-$73:19:13.01 & 0.0544 & 18.2 & June 4 2018 & 3400 & $-4.47$\\
3FHL J2240.3$-$5240 & SUMSS J224017$-$524111 & 22:40:17.64 & $-$52:41:13.07 & 0.0118 & 16.7 & June 4 2018 & 1950 & $-5.84$\\
3FHL J2321.8$-$6437 & PMN J2321$-$6438 & 23:21:42.17 & $-$64:38:06.90 & 0.02 & 17.4 & June 4 2018 & 2800 & $-0.06$\\
3FHL J2339.2$-$7404 & 1RXS J233919.8$-$740439 & 23:39:20.88 & $-$74:04:36.12 & 0.0262 & 16.1 & June 4 2018 & 1500 & $-0.65$\\
\hline 

\hline
\hline
\end{tabular}}\caption{List of sources and their properties sorted in the order of increasing R.A. (Right ascension) values. (1): 3FHL catalog \citep{ajello17} name for the source. (2): optical, IR, X-ray or radio counterpart of the source. (3) Right ascension. (4) Declination. (5) $E(B-V)$ value obtained using the measurements of \citet{schlafly11} and the NASA/IPAC Infrared Science Archive online tool. (6) V band magnitude. (7) Date of observation. (8) Exposure time (in seconds).(9)Slope of continuum fit obtained from the observed fits file }
\label{tab:sample}
\end{table*}
\endgroup

\begingroup
\renewcommand*{\arraystretch}{1.8}
\begin{table*}
\centering
\scalebox{0.65}{
\begin{tabular}{cccccc}
\hline
\hline
Source & S/N & Spectral line & Observed $\lambda$ (\r{A}) & line type & redshift\\
 & & Rest frame $\lambda$ (\r{A})& & &\\
\hline
3FHL J0002.1$-$6728 & 41.4 & &  &  & \\
3FHL J0935.2$-$1735 & 51.5 & &  &  & \\
3FHL J0936.4$-$2109 & 27.2  & Mg II(2797) & 6176 & absorption & $>1.197^*$\\
 & & Mg II(2803) & 6160 & absorption & \\
3FHL J1030.6$-$2029 & 29.3 & Mg II(2797) & 5579 & absorption & $>0.995$\\
 & & Mg II(2803) & 5591 & absorption & \\
3FHL J1042.8$+$0055 & 46.6  & &  &  & \\
3FHL J1130.5$-$7801 & 72.2  & &  &  & \\
3FHL J1155.5$-$3418 & 42.7 & Mg II(2797) & 5174 & absorption & $>0.849$\\
 & & Mg II(2803) & 5185 & absorption & \\
3FHL J1212.1$-$2328 & 102.8 & O III(5007) & 8345 & emission & $0.666$\\ 
3FHL J1223.5$-$3033 & 46.5 & Mg II(2797) & 5245 & absorption & $>0.875^*$\\
 & & Mg II(2803) & 5256 & absorption & \\
3FHL J1229.7$-$5304 & 78.6  & &  &  & \\
3FHL J1315.9$-$0732 & 60.8  & &  &  & \\
3FHL J1433.5$-$7304 & 64.9 & G-band(4304) & 5165 & absorption & $0.200$\\
 & & Mg I(5175) & 6209 & absorption & \\
 & & Ca+Fe(5269) & 6340 & absorption & \\
 & & Na (5895) & 7074 & absorption & \\
 & & H$\_${$\alpha$}(6562) & 7876 & absorption & \\
3FHL J1439.4$-$2524 & 82.7  & Mg I(5175) & 6008 & absorption & $0.16$\\
 & & Ca+Fe(5269) & 6115 & absorption & \\
 & & NaD(5892) & 6835 & absorption & \\  
3FHL  J1605.0$-$1140 & 17.2 & O II(3727) & 6801 & emission & $0.358^*$\\
 & & (or) O III(5007) & 6801 & emission & $0.824^*$\\
3FHL J1612.3$-$3100 & 75.4  & &  &  & \\
3FHL J1640.1$+$0629 & 83.1 & & &  & \\
3FHL J1842.4$-$5841 & 32.7 &  & &  & \\
3FHL J1924.2$-$1548 & 64.4  &  & &  & \\
%3FHL J1944.4$-$4523 & 122.6 &  & &  & \\
3FHL J2034.9$-$4200 & 33.4 &  & &  & \\
3FHL J2041.7$-$7319 & 70.1 & & &  & \\
3FHL J2240.3$-$5240 & 71.2 & & &  & \\
3FHL J2321.8$-$6437 & 33.7 & & &  & \\
3FHL J2339.2$-$7404 & 45.5 & & &  & \\

\hline

\hline
\hline
\end{tabular}}\caption{Results obtained from spectral analysis discussed in Section~\ref{spectral}. The redshift measurement values marked with a $^*$ are tentative $z$ measurements.}
\label{tab:redshift}
\end{table*}
\endgroup

\begin{figure*}
\begin{minipage}[b]{.5\textwidth}
  \centering
  \includegraphics[width=0.9\textwidth]{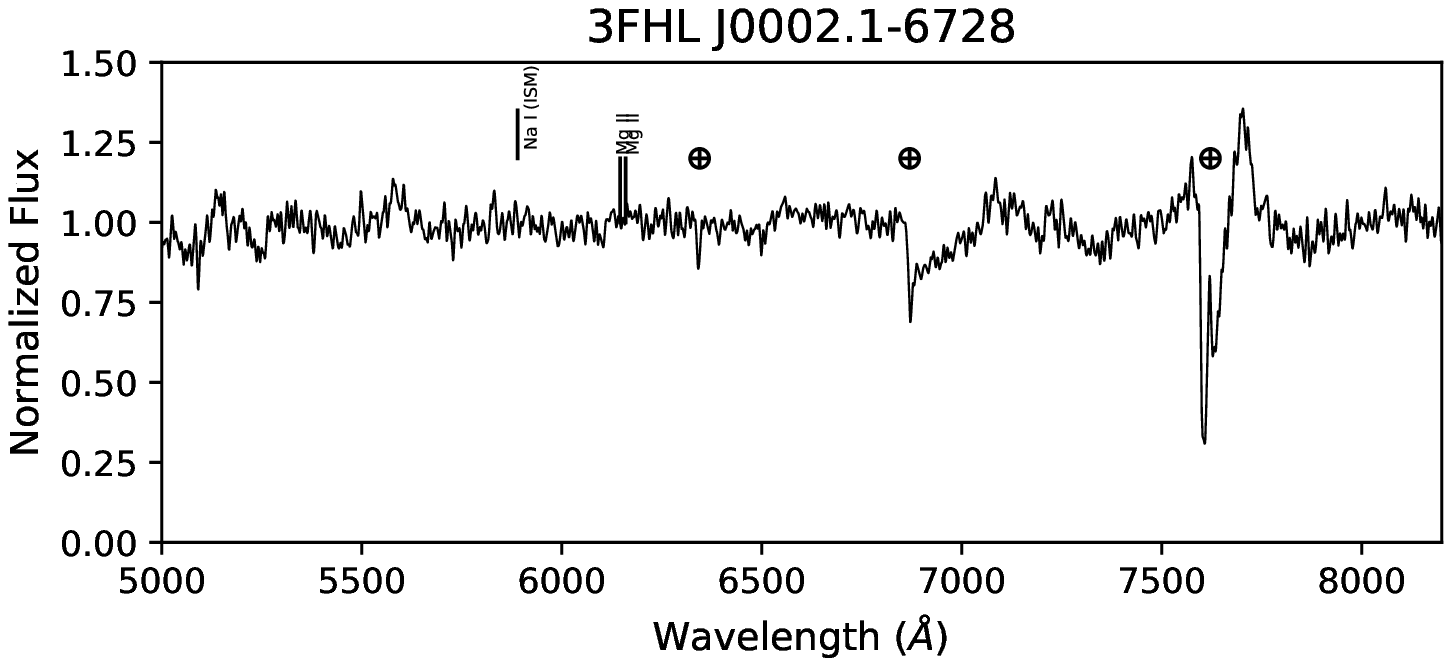}
  \end{minipage}
\begin{minipage}[b]{.5\textwidth}
  \centering
  \includegraphics[width=0.9\textwidth]{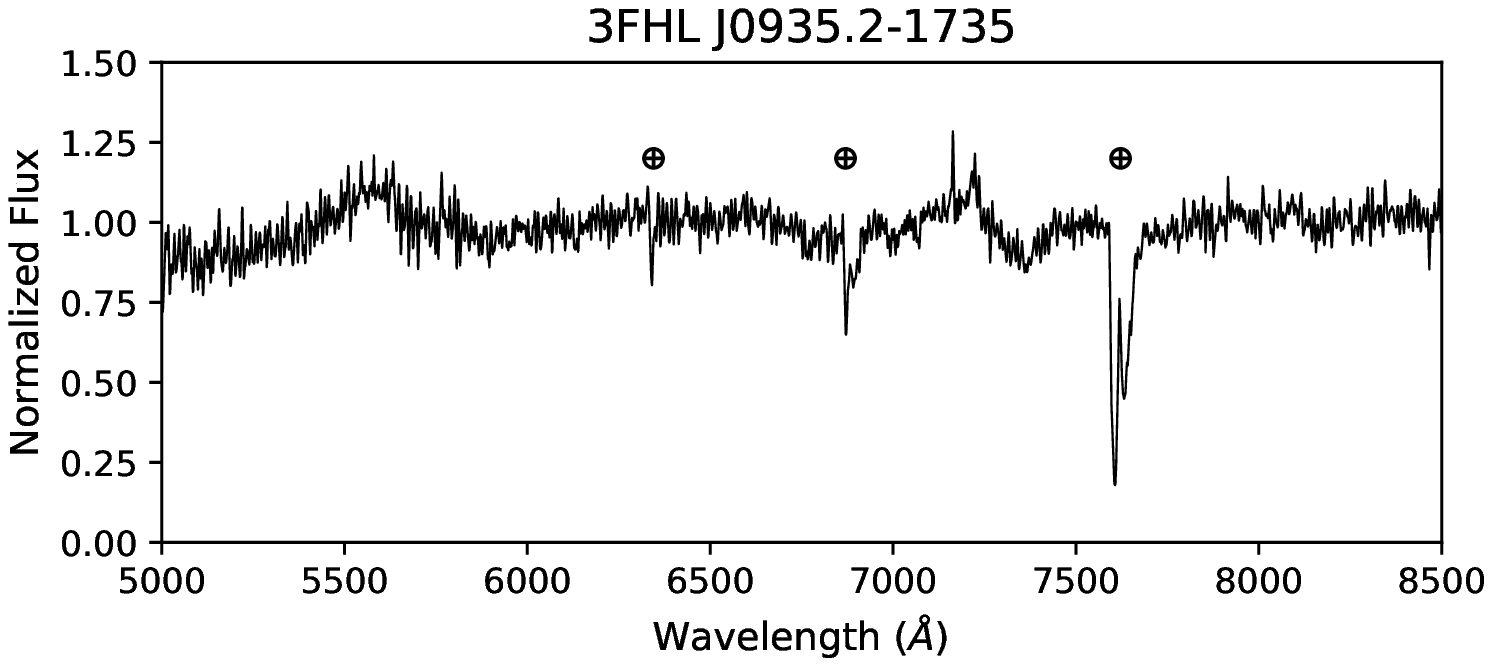}
  \end{minipage}
\begin{minipage}[b]{.5\textwidth}
  \centering
  \includegraphics[width=0.9\textwidth]{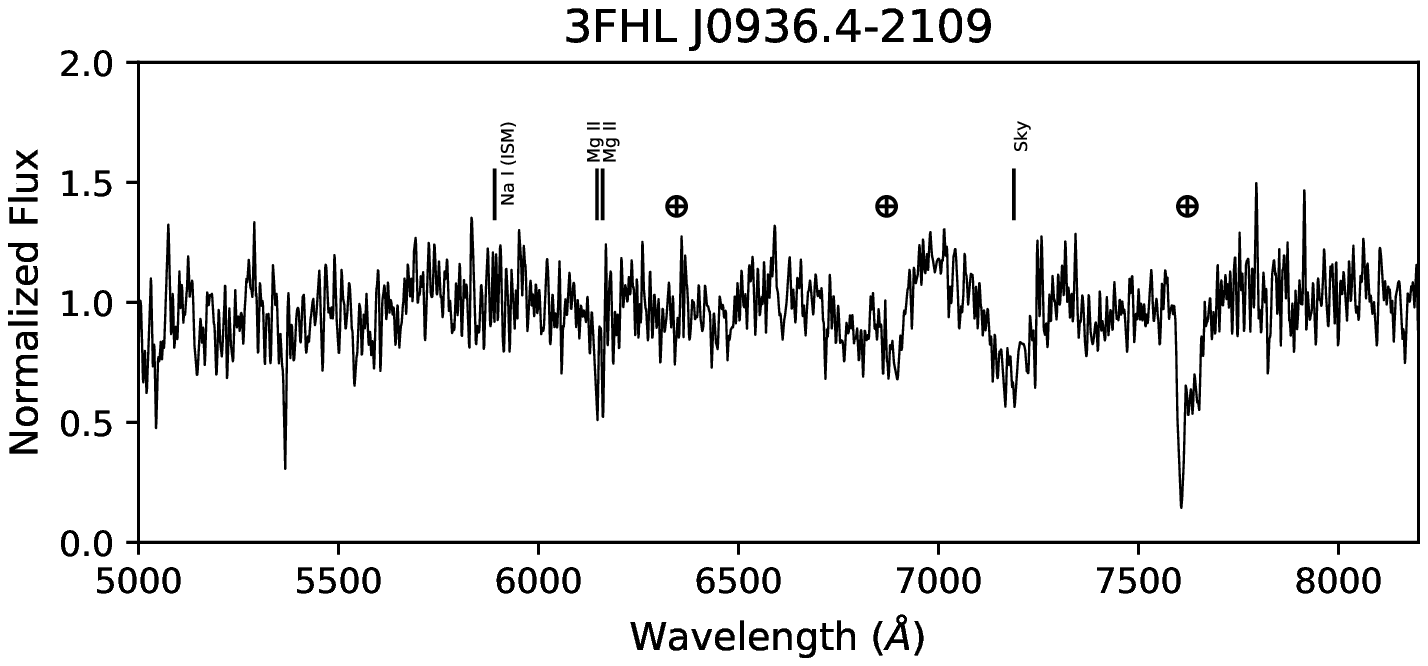}
  \end{minipage}
\begin{minipage}[b]{.5\textwidth}
  \centering
  \includegraphics[width=0.9\textwidth]{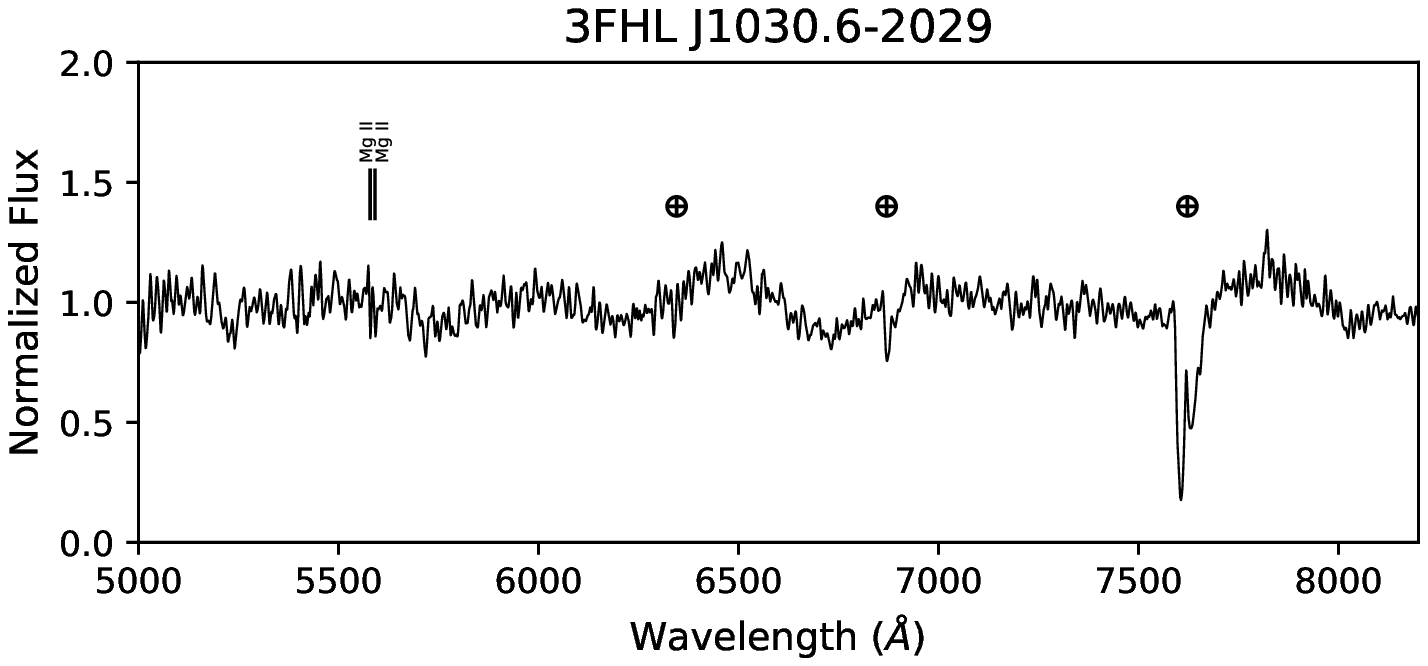}
  \end{minipage}
    \vspace{0.5cm}
\begin{minipage}[b]{.5\textwidth}
  \centering
  \includegraphics[width=0.9\textwidth]{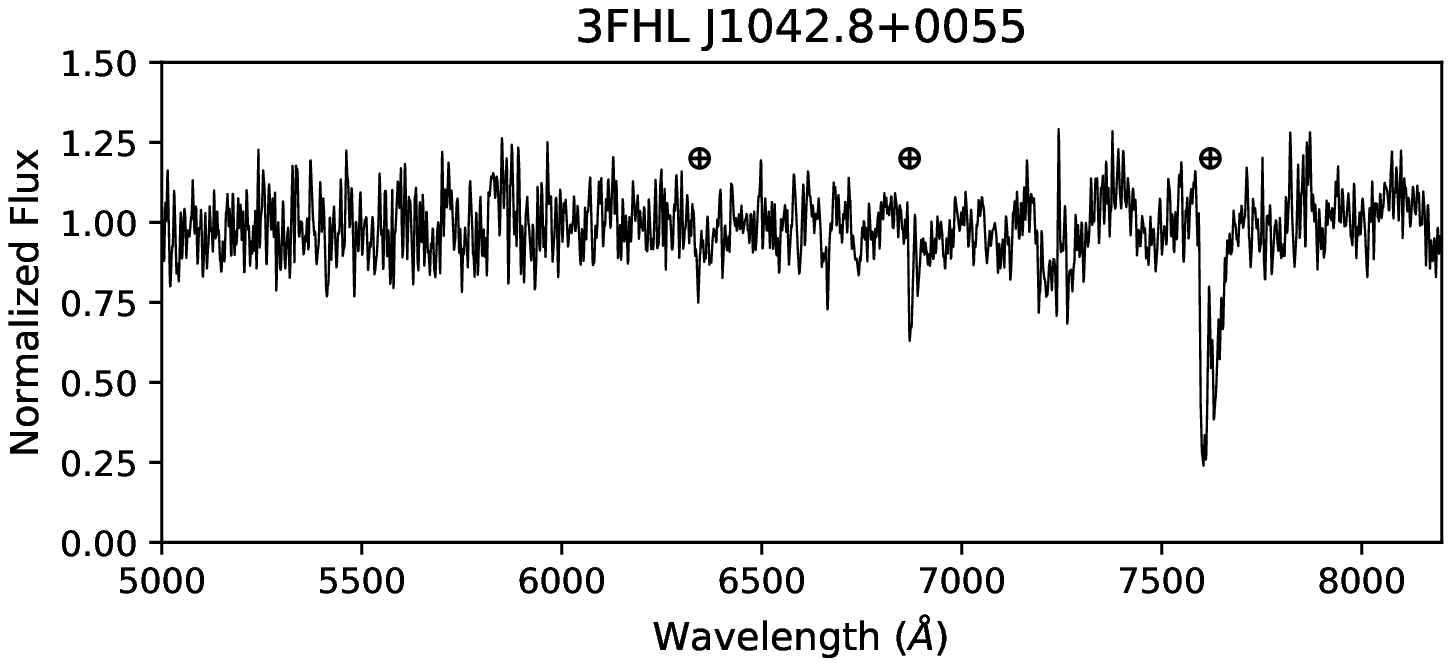}
  \end{minipage}
  \begin{minipage}[b]{.5\textwidth}
  \centering
  \includegraphics[width=0.9\textwidth]{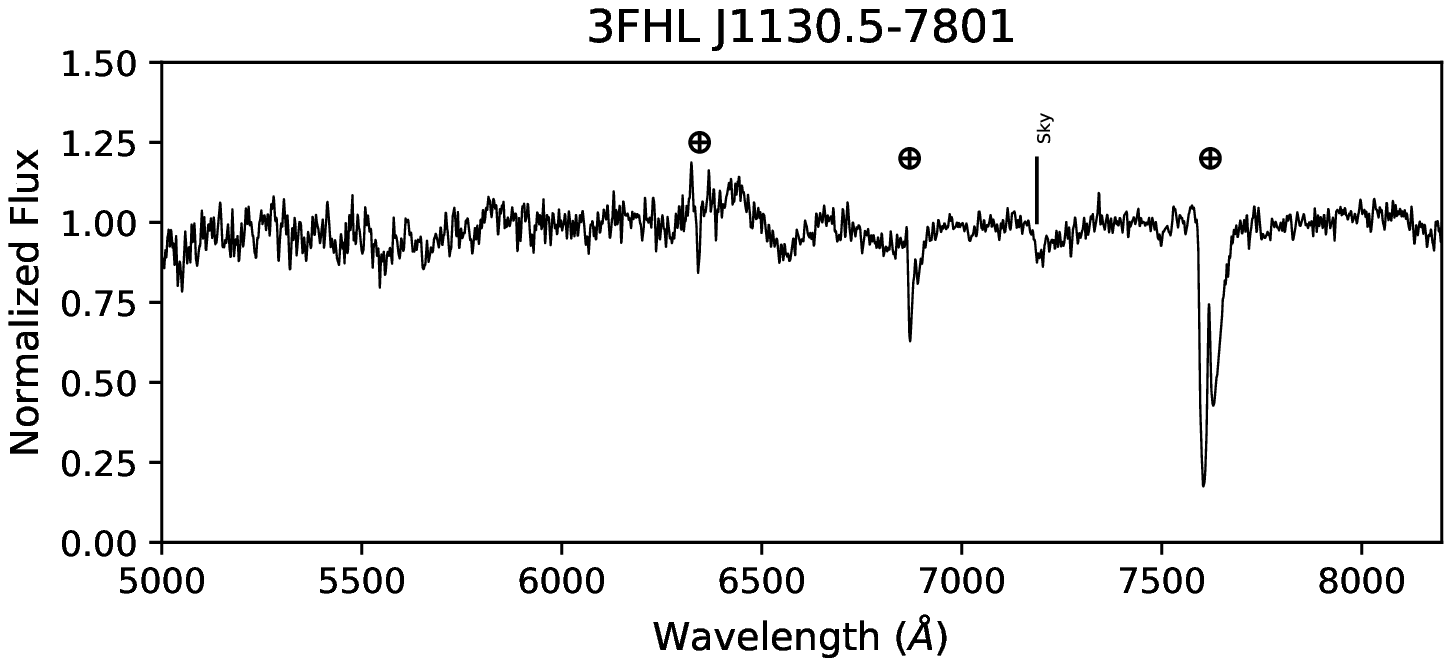}
  \end{minipage}
  \begin{minipage}[b]{.5\textwidth}
  \centering
  \includegraphics[width=0.9\textwidth]{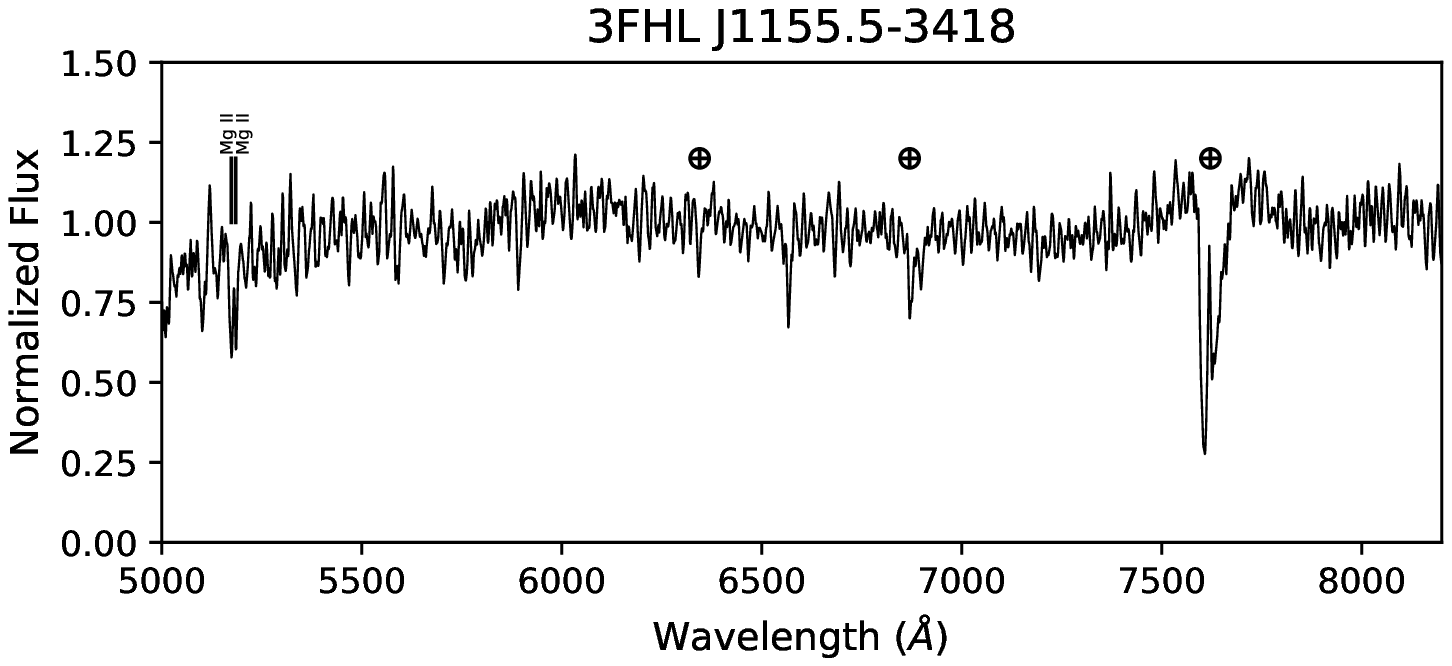}
  \end{minipage}
\begin{minipage}[b]{.5\textwidth}
  \centering
  \includegraphics[width=0.9\textwidth]{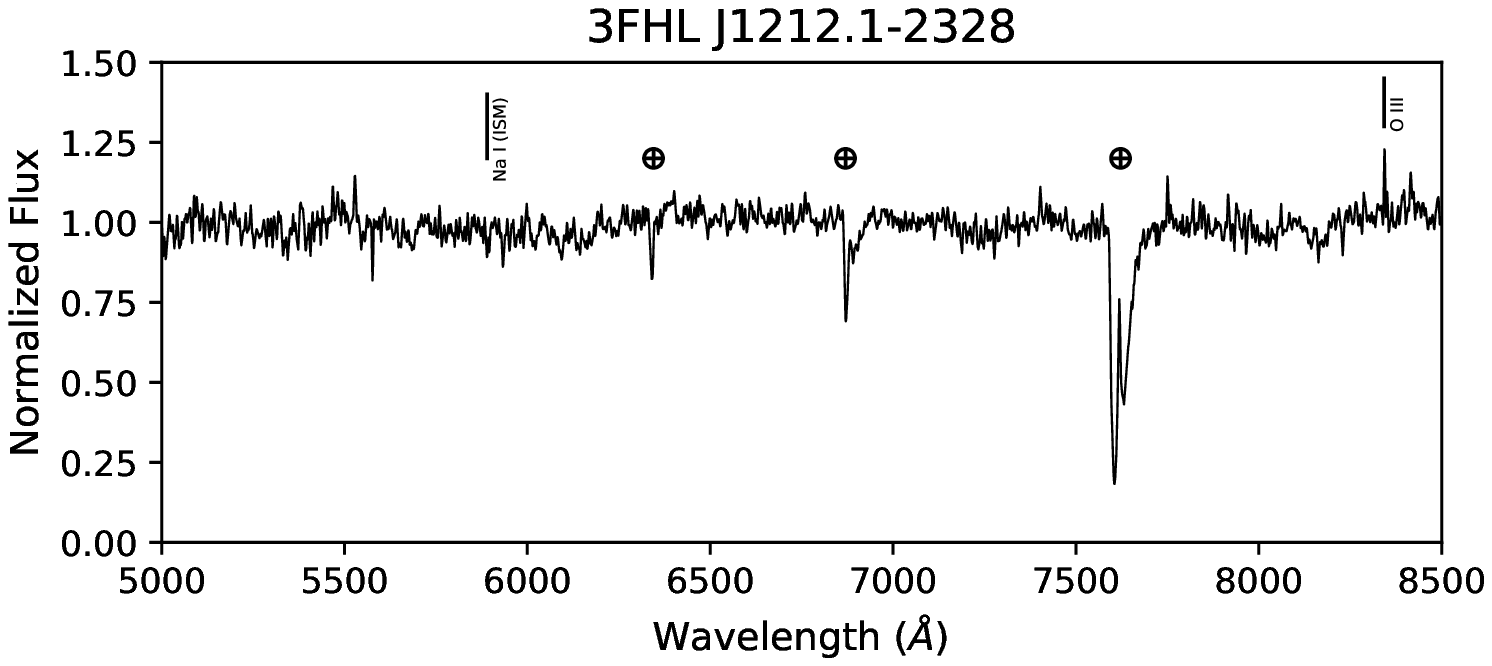}
  \end{minipage}
\begin{minipage}[b]{.5\textwidth}
  \centering
  \includegraphics[width=0.9\textwidth]{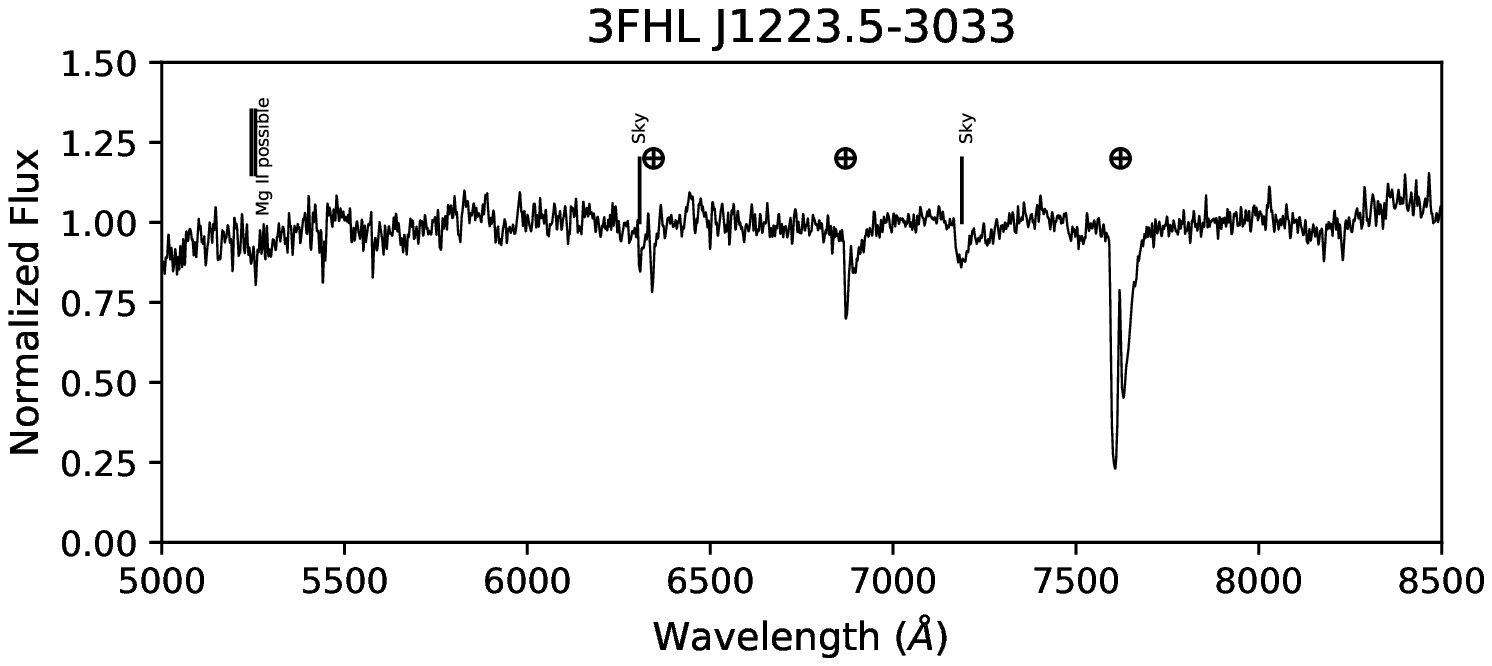}
  \end{minipage}
\begin{minipage}[b]{.5\textwidth}
  \centering
  \includegraphics[width=0.9\textwidth]{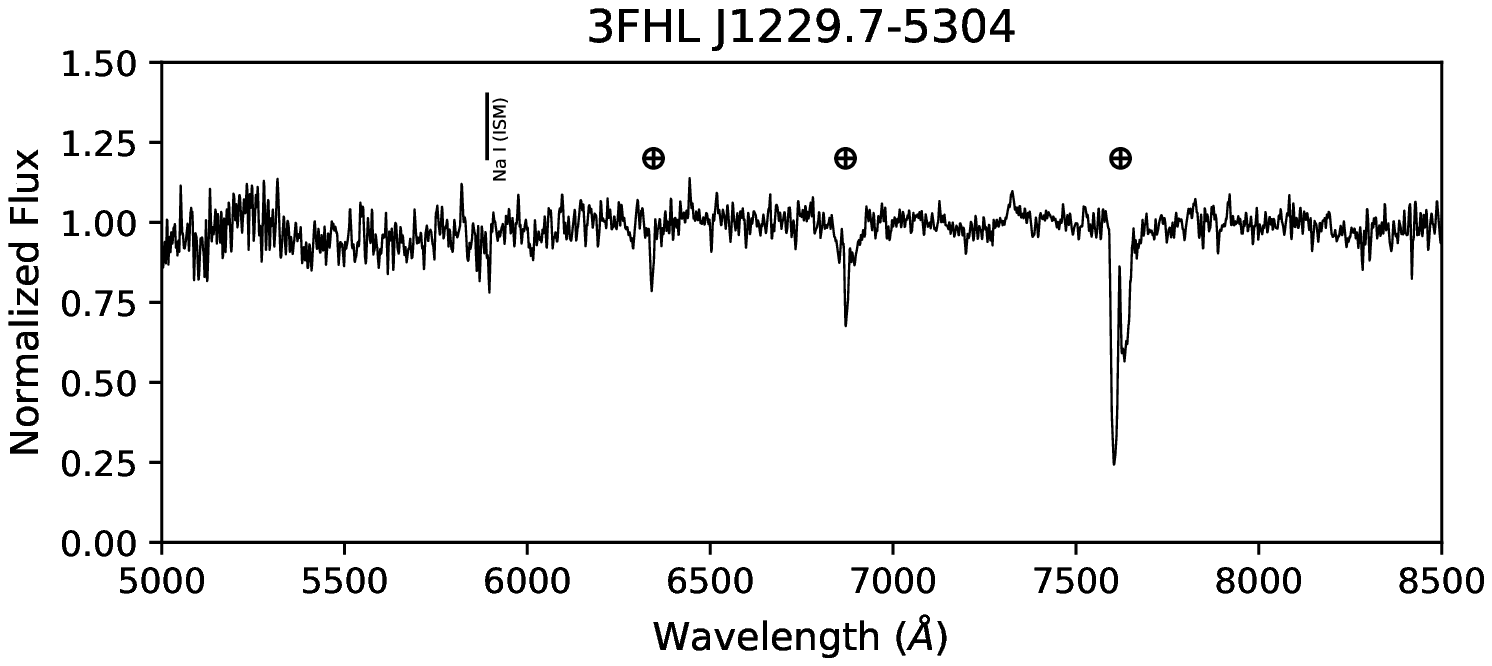}
  \end{minipage}
\caption{Optical spectra of the observed candidates after performing flux calibration and dereddening. The bottom panel displays the normalized spectra where the atmospheric features are denoted by $\otimes$ while the absorption or emission features are labeled as per the lines they signify.}
\label{fig:spec}
\end{figure*}

\begin{figure*}
\setcounter{figure}{1}

    \vspace{0.5cm}
  \begin{minipage}[b]{.5\textwidth}
  \centering
  \includegraphics[width=0.9\textwidth]{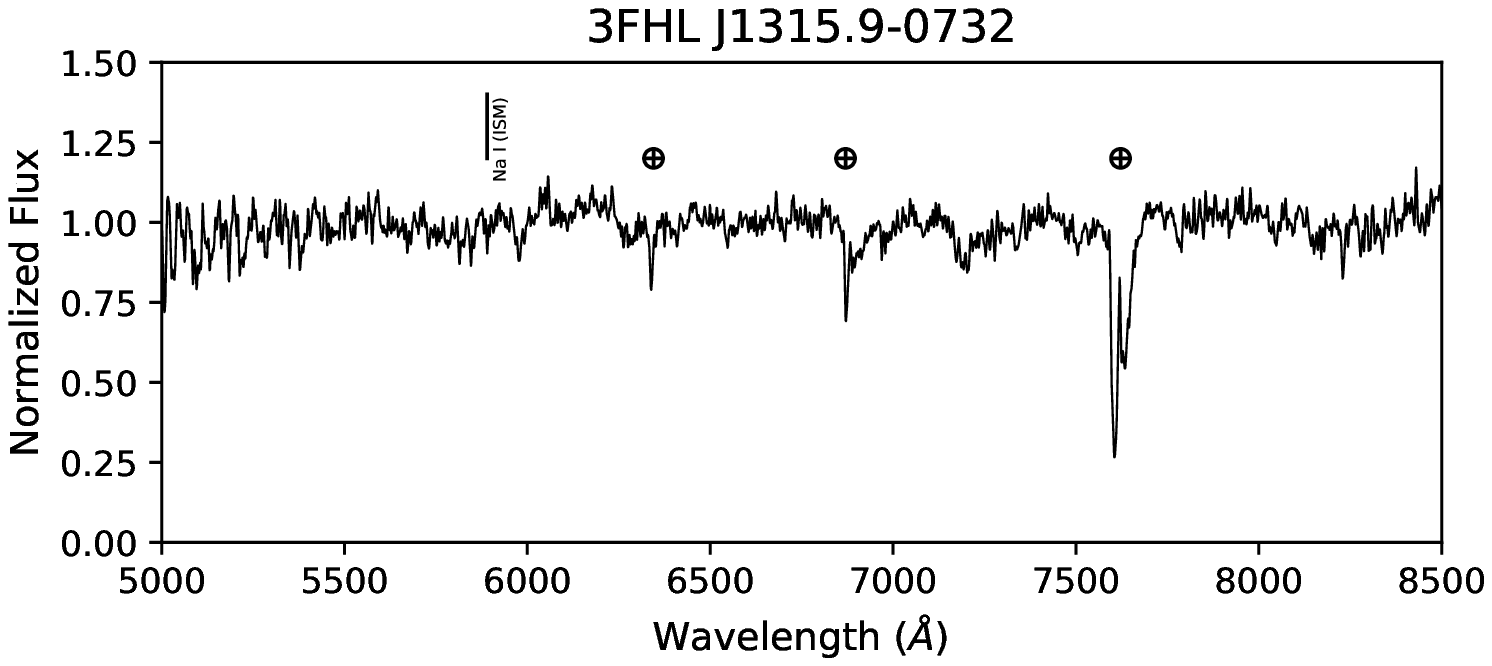}
  \end{minipage}
  \begin{minipage}[b]{.5\textwidth}
  \centering
  \includegraphics[width=0.9\textwidth]{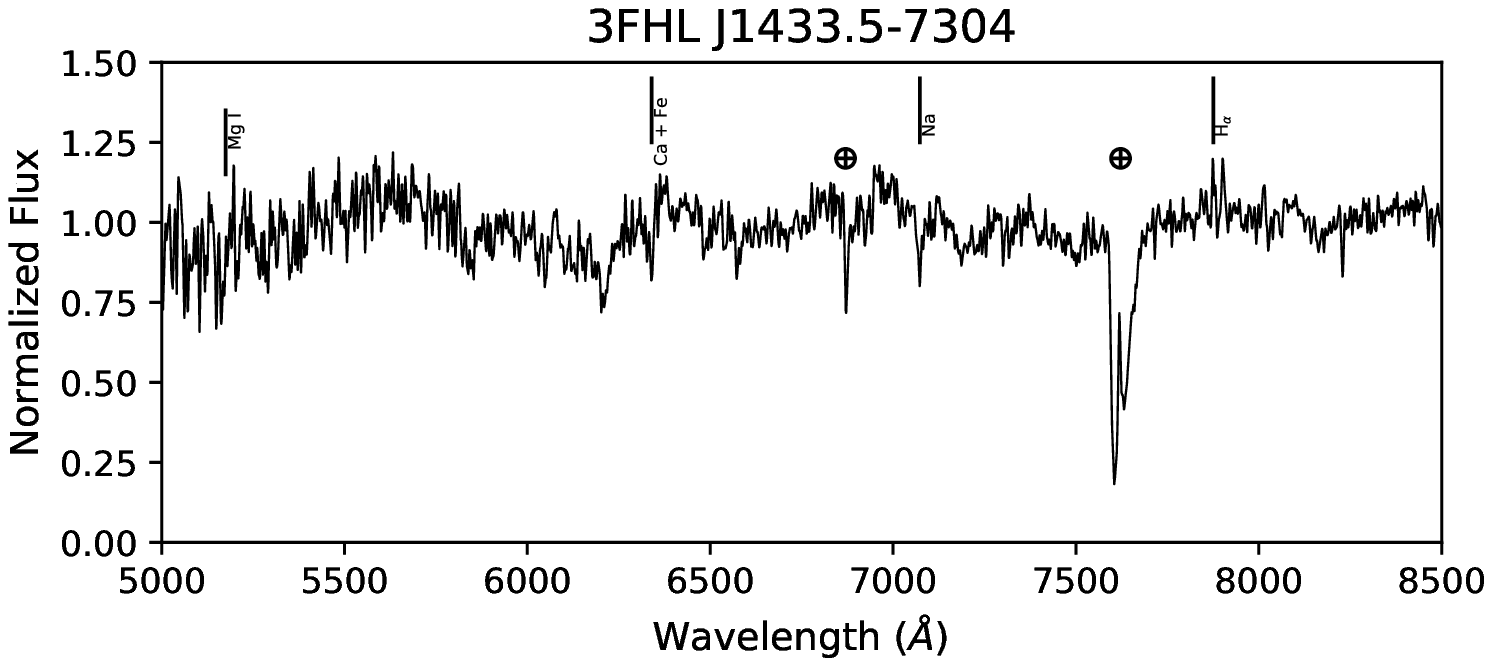}
  \end{minipage}
  \begin{minipage}[b]{.5\textwidth}
  \centering
  \includegraphics[width=0.9\textwidth]{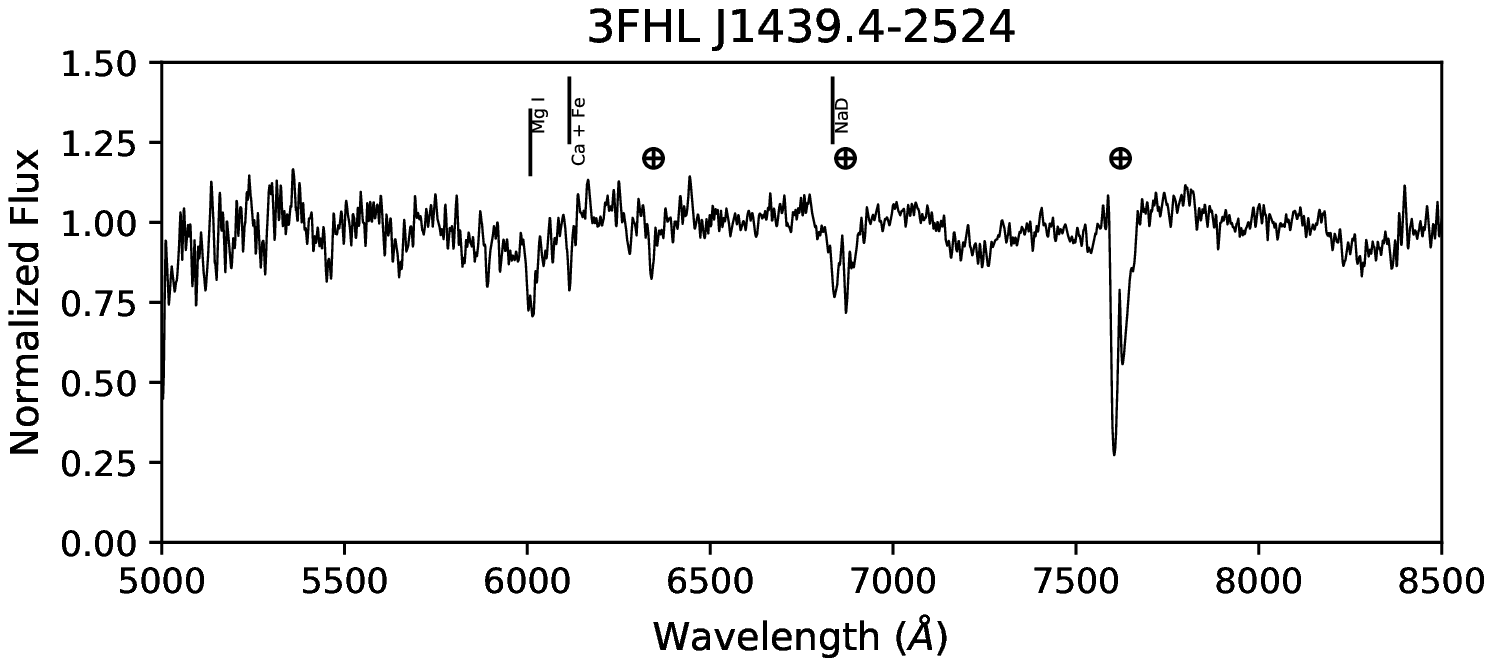}
  \end{minipage}
\begin{minipage}[b]{.5\textwidth}
  \centering
  \includegraphics[width=0.9\textwidth]{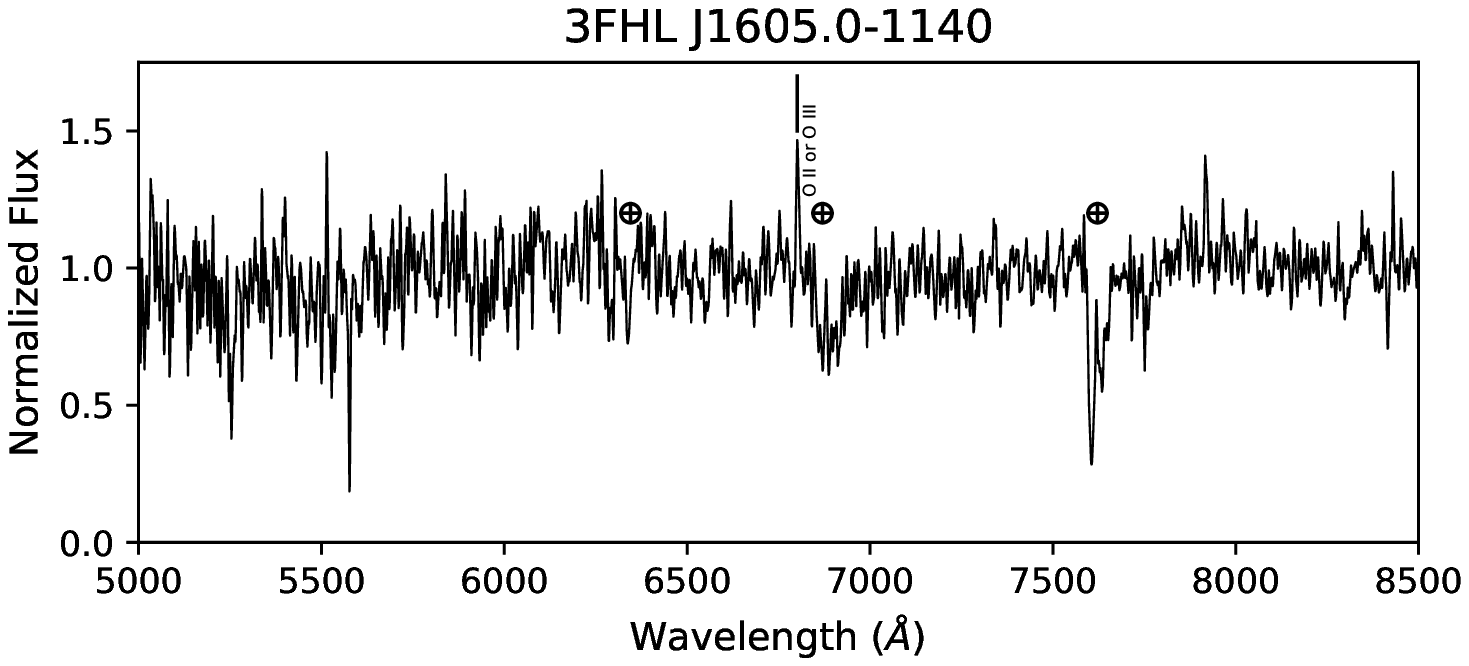}
  \end{minipage}
\begin{minipage}[b]{.5\textwidth}
  \centering
  \includegraphics[width=0.9\textwidth]{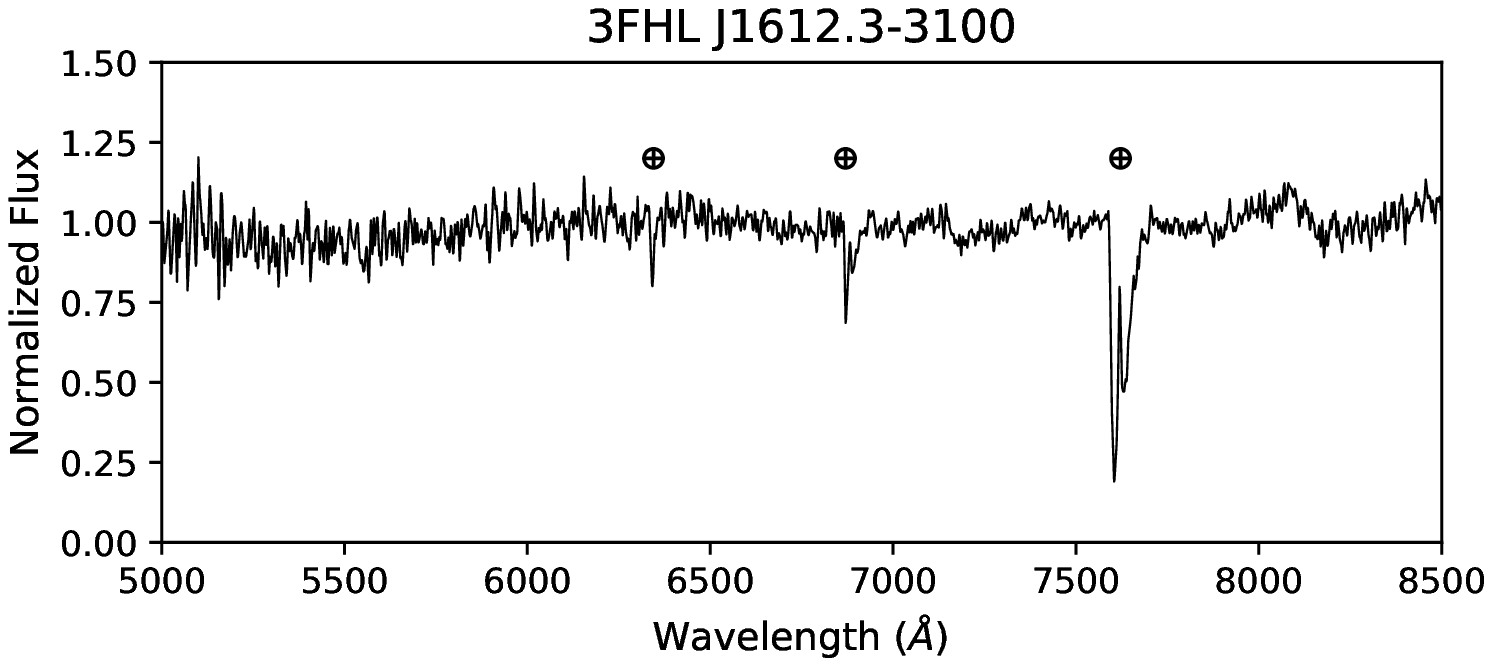}
  \end{minipage}
\begin{minipage}[b]{.5\textwidth}
  \centering
  \includegraphics[width=0.9\textwidth]{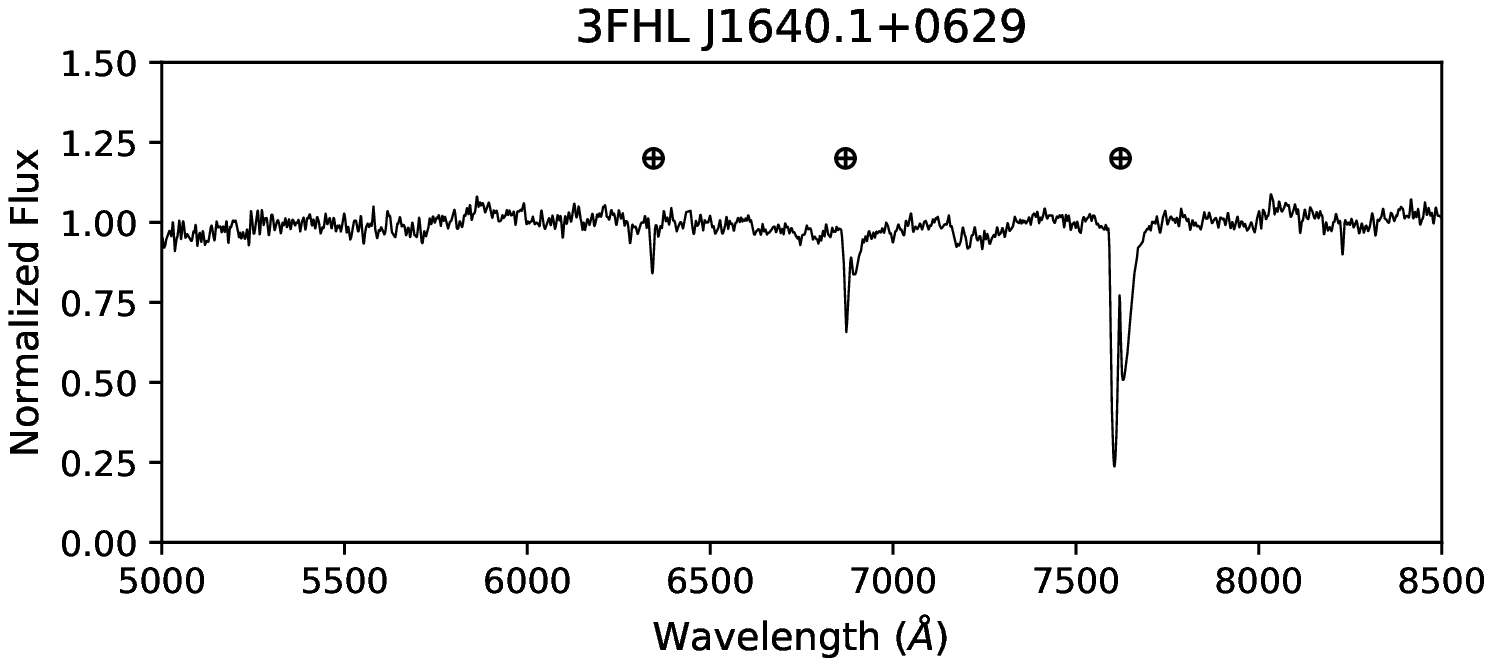}
  \end{minipage}
    \vspace{0.5cm}
\begin{minipage}[b]{.5\textwidth}
  \centering
  \includegraphics[width=0.9\textwidth]{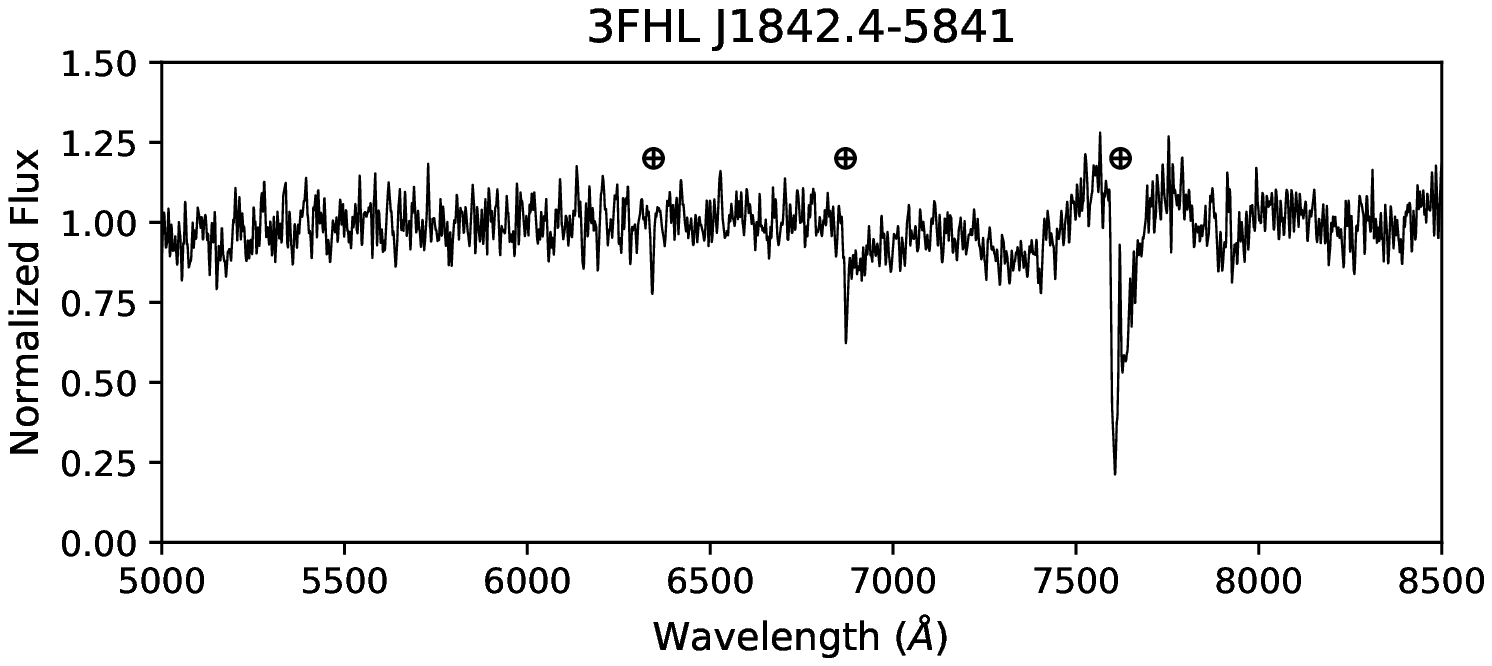}
  \end{minipage}
  \begin{minipage}[b]{.5\textwidth}
  \centering
  \includegraphics[width=0.9\textwidth]{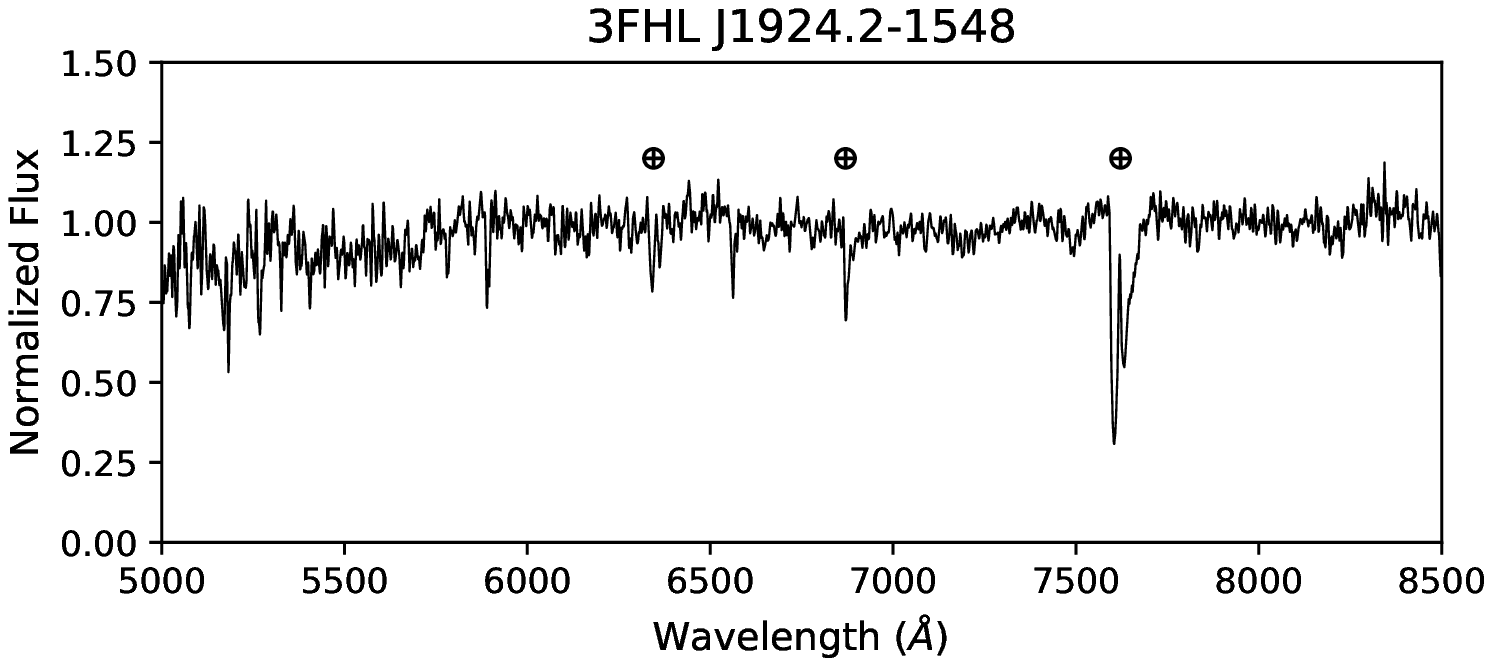}
  \end{minipage}
  \begin{minipage}[b]{.5\textwidth}
  \centering
  \includegraphics[width=0.9\textwidth]{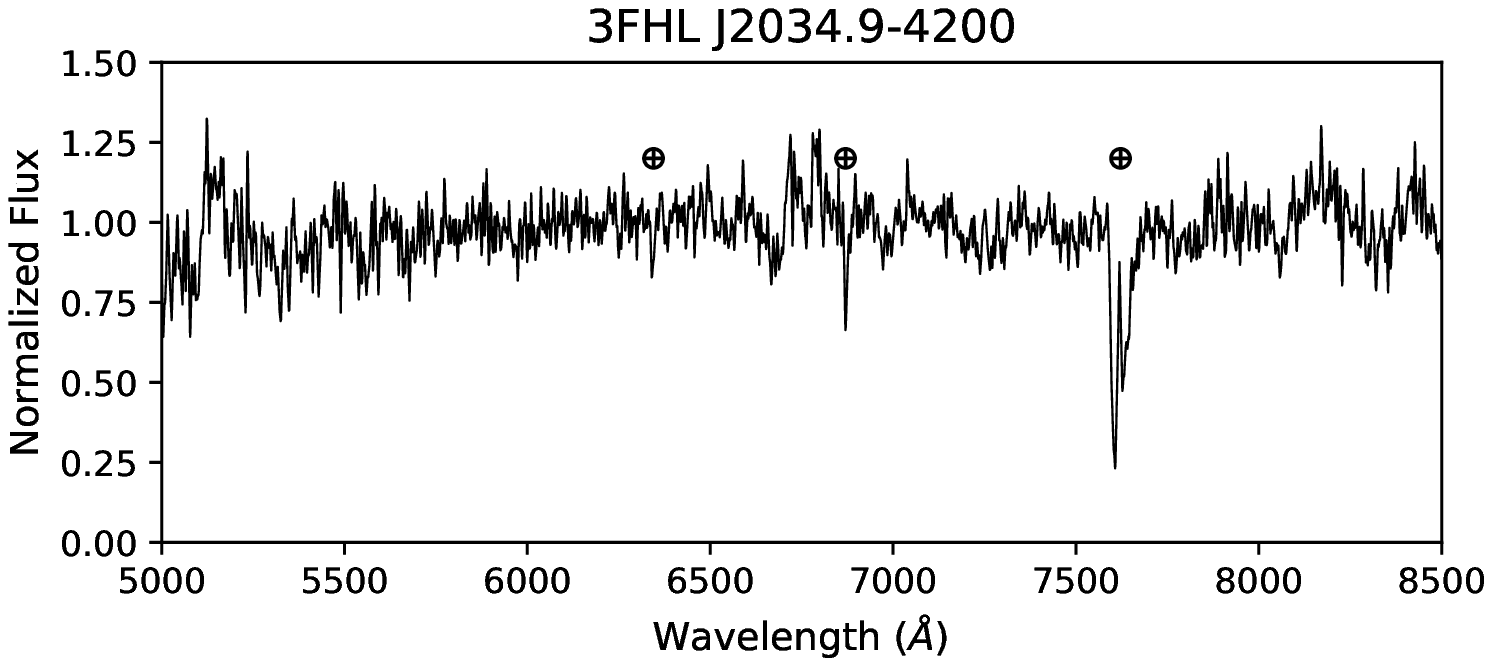}
  \end{minipage}
\begin{minipage}[b]{.5\textwidth}
  \centering
  \includegraphics[width=0.9\textwidth]{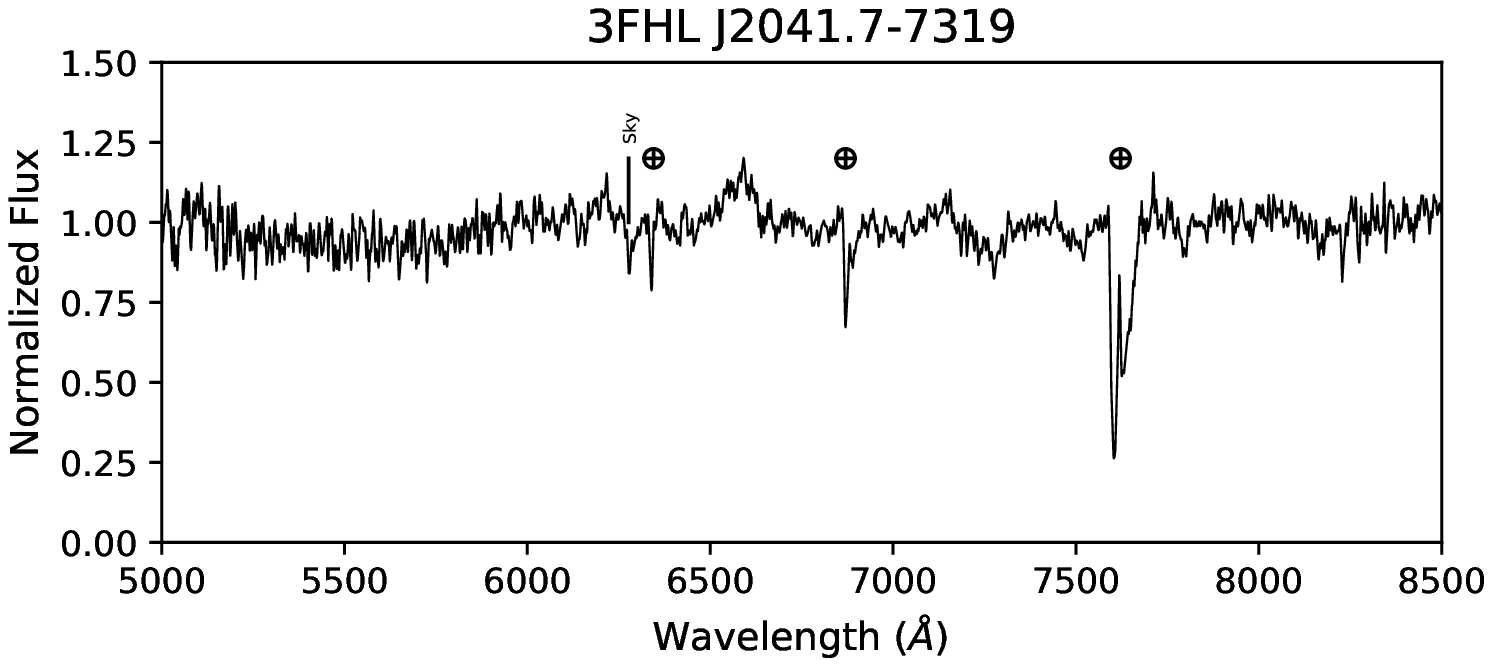}
  \end{minipage}

\caption{Continued from Fig~\ref{fig:spec}}
\end{figure*}

\begin{figure*}
\setcounter{figure}{1}

\begin{minipage}[b]{.5\textwidth}
  \centering
  \includegraphics[width=0.9\textwidth]{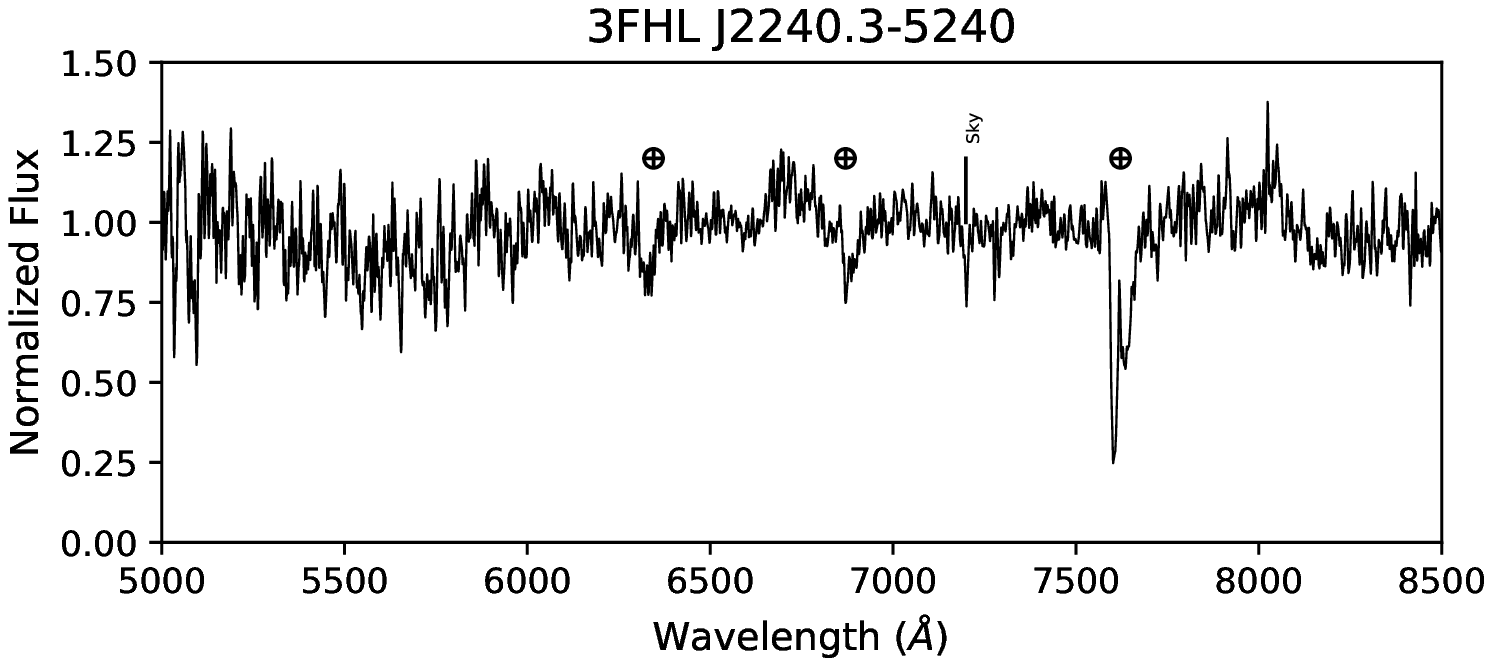}
  \end{minipage}
\begin{minipage}[b]{.5\textwidth}
  \centering
  \includegraphics[width=0.9\textwidth]{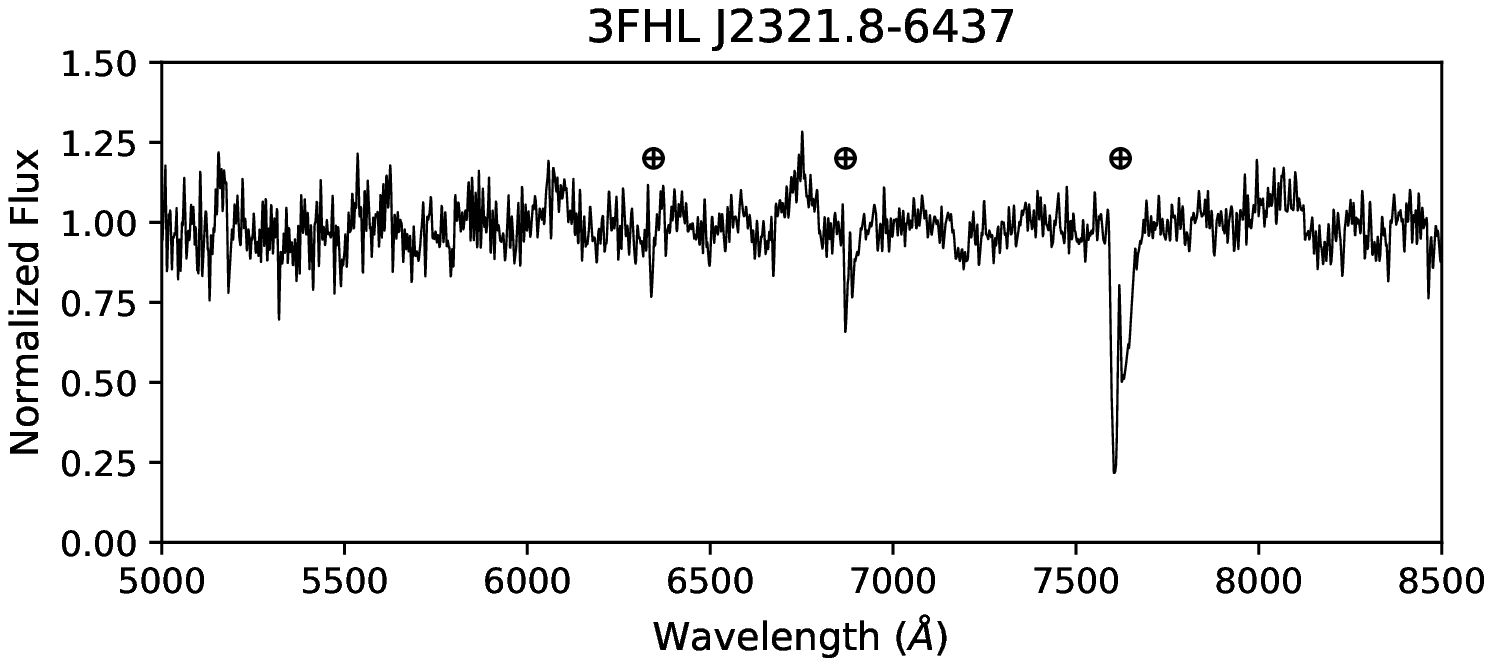}
  \end{minipage}
    \vspace{0.5cm}
\begin{minipage}[b]{.5\textwidth}
  \centering
  \includegraphics[width=0.9\textwidth]{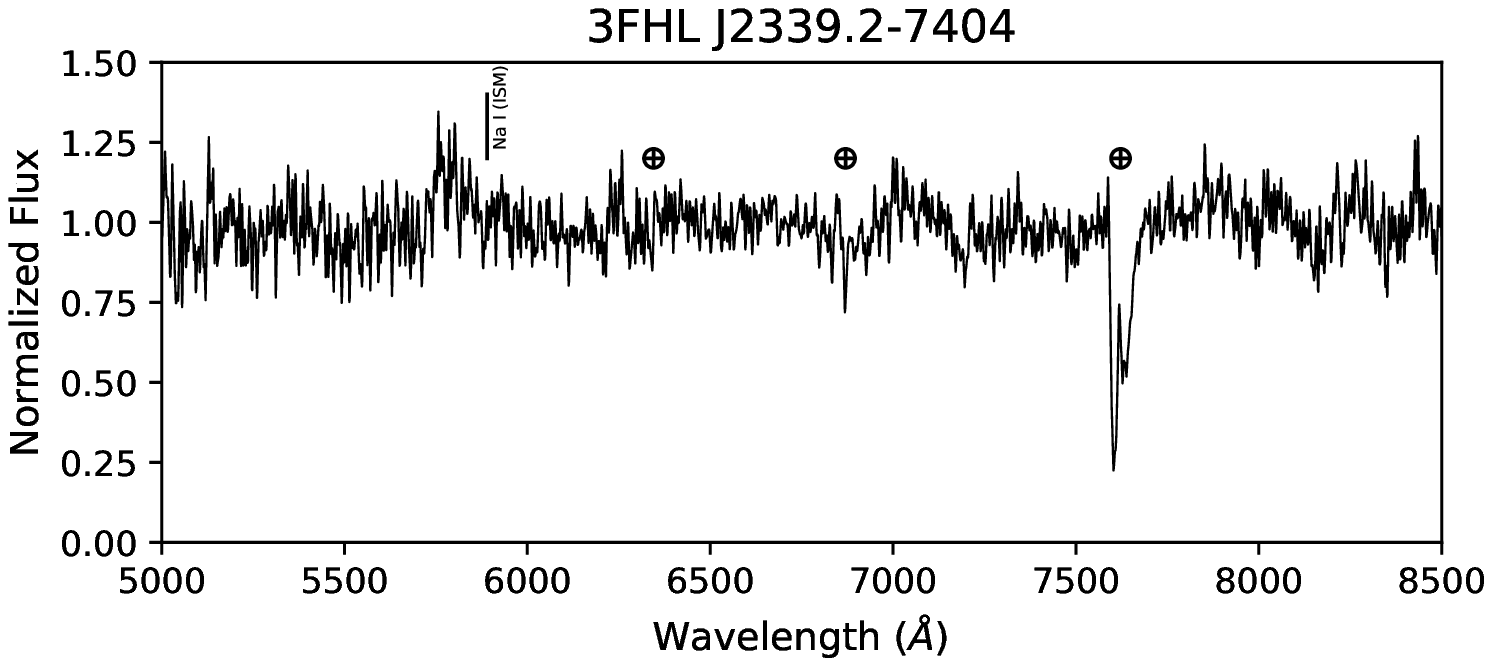}
  \end{minipage}
  \begin{minipage}[b]{.5\textwidth}
  \centering
  \end{minipage}
\caption{Continued from Fig~\ref{fig:spec}}
\end{figure*}

\begin{figure*}

\begin{minipage}[b]{.5\textwidth}
  \centering
  \includegraphics[width=0.9\textwidth]{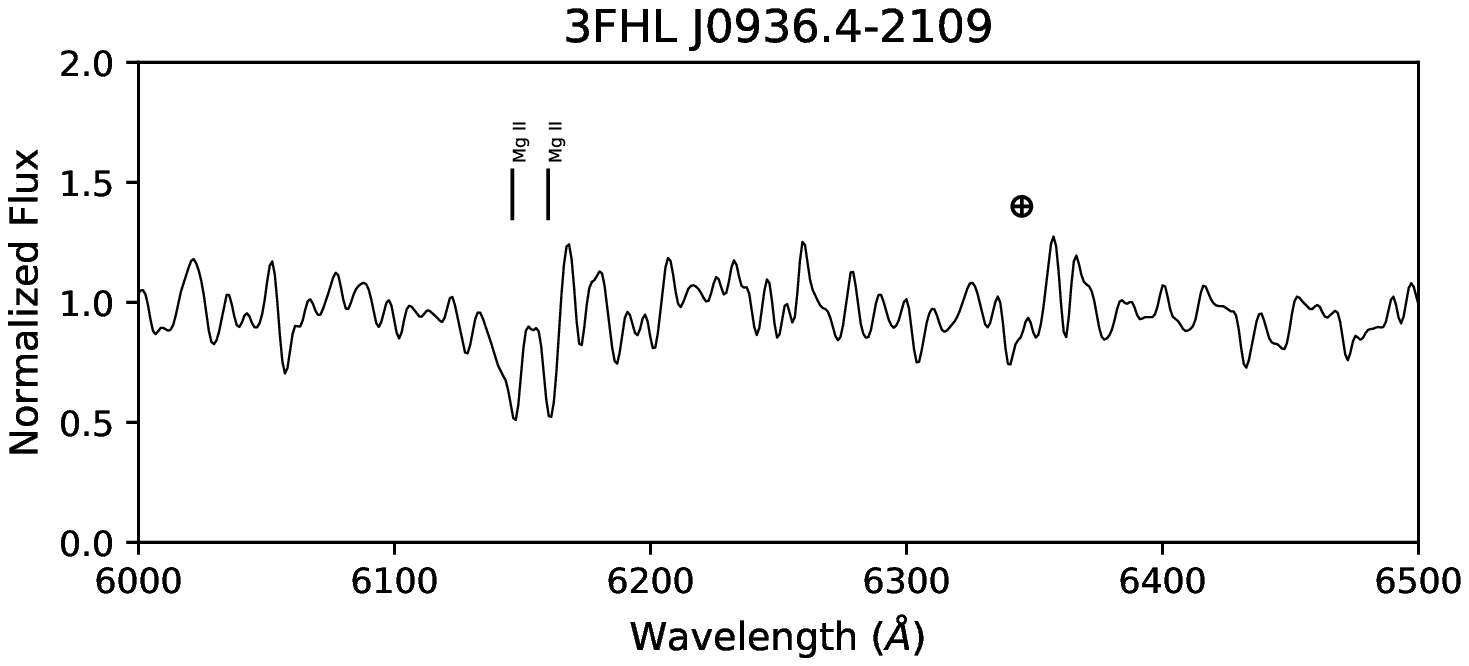}
  \end{minipage}
\begin{minipage}[b]{.5\textwidth}
  \centering
  \includegraphics[width=0.9\textwidth]{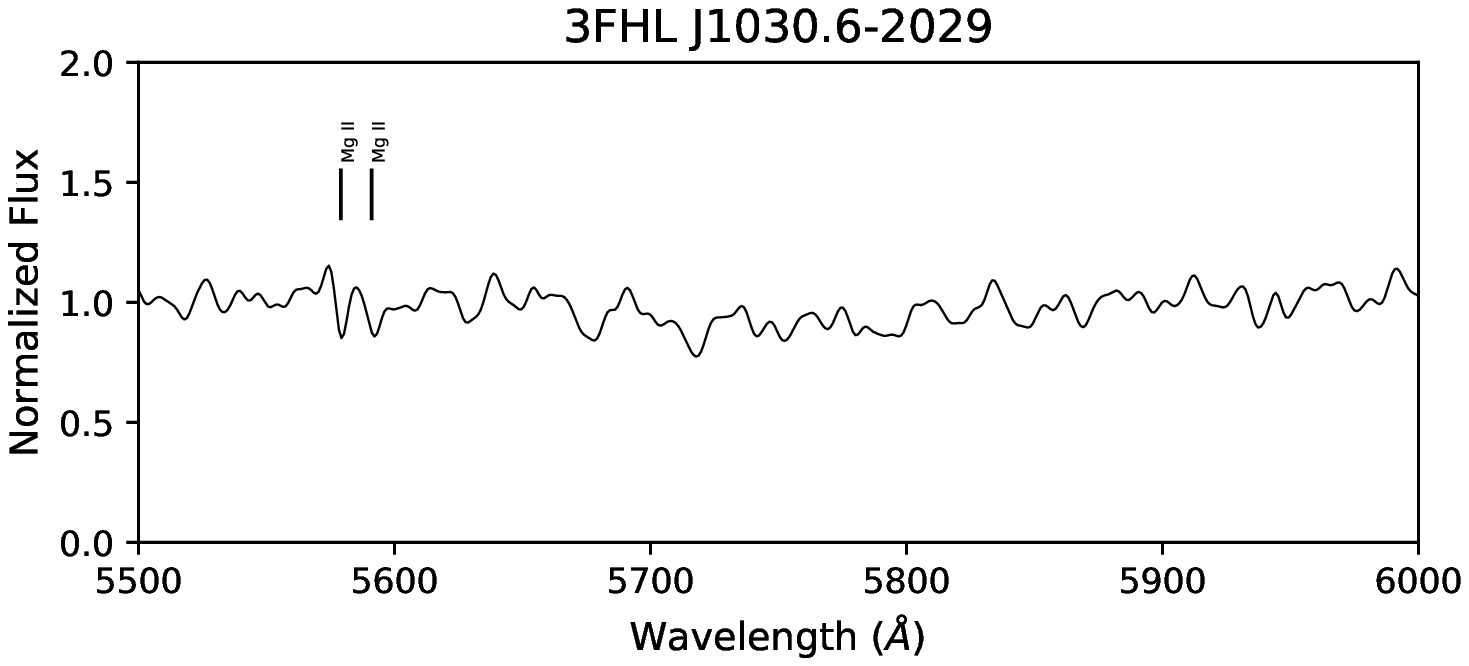}
  \end{minipage}
    \vspace{0.5cm}
\begin{minipage}[b]{.5\textwidth}
  \centering
  \includegraphics[width=0.9\textwidth]{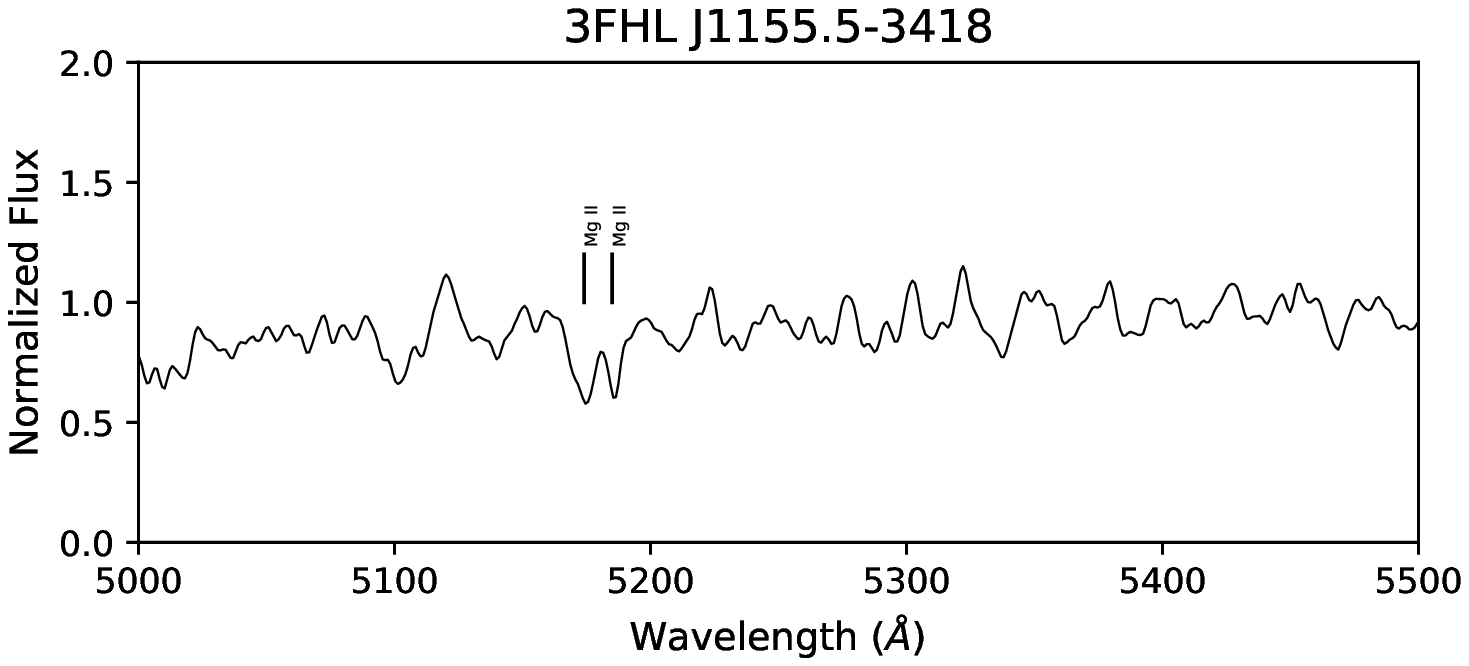}
  \end{minipage}
  \begin{minipage}[b]{.5\textwidth}
  \centering
  \includegraphics[width=0.9\textwidth]{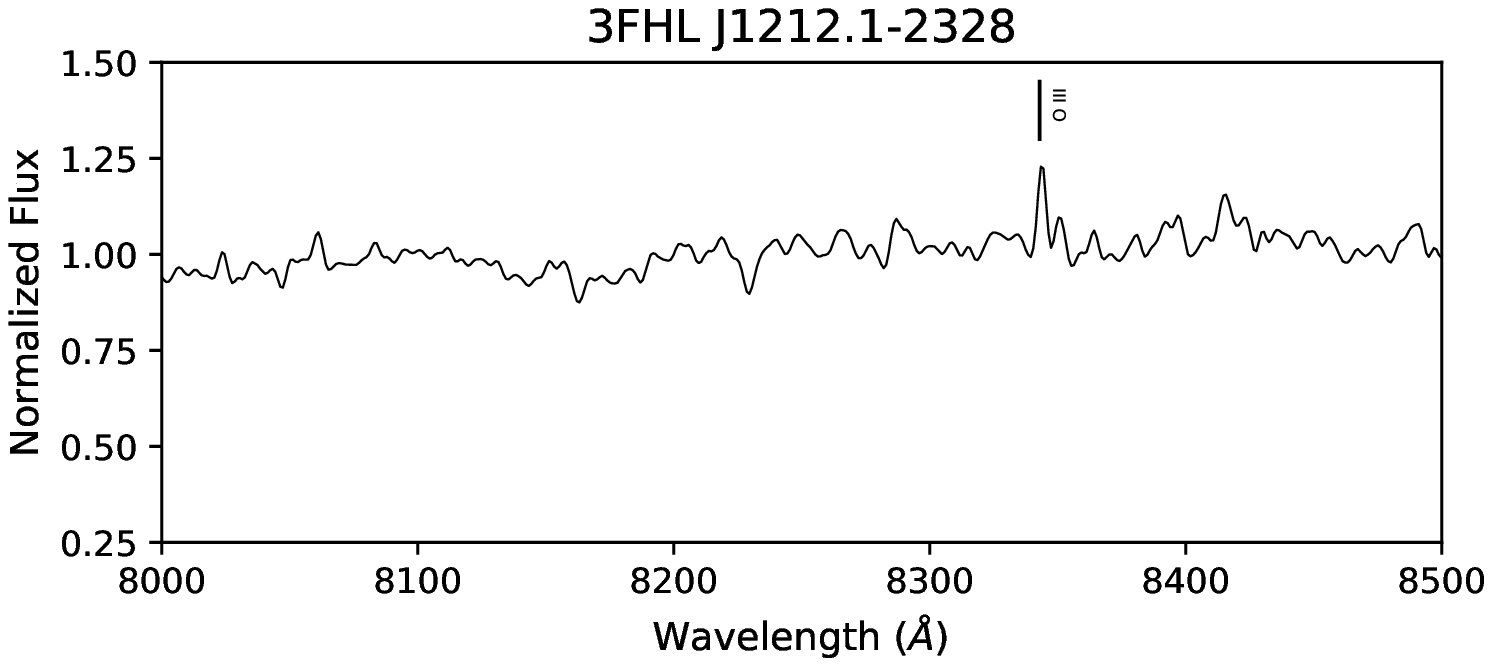}
  \end{minipage}
  \begin{minipage}[b]{.5\textwidth}
  \centering
  \includegraphics[width=0.9\textwidth]{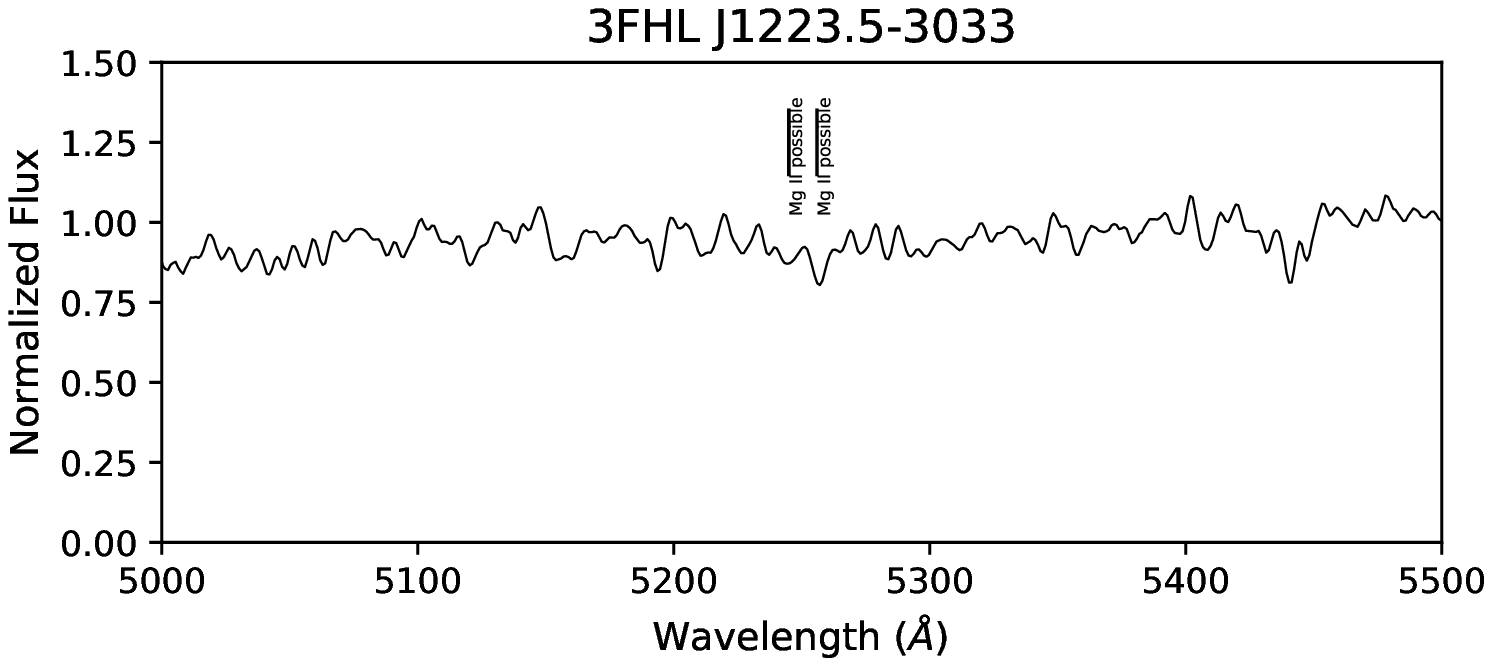}
  \end{minipage}
  \begin{minipage}[b]{.5\textwidth}
  \centering
  \includegraphics[width=0.9\textwidth]{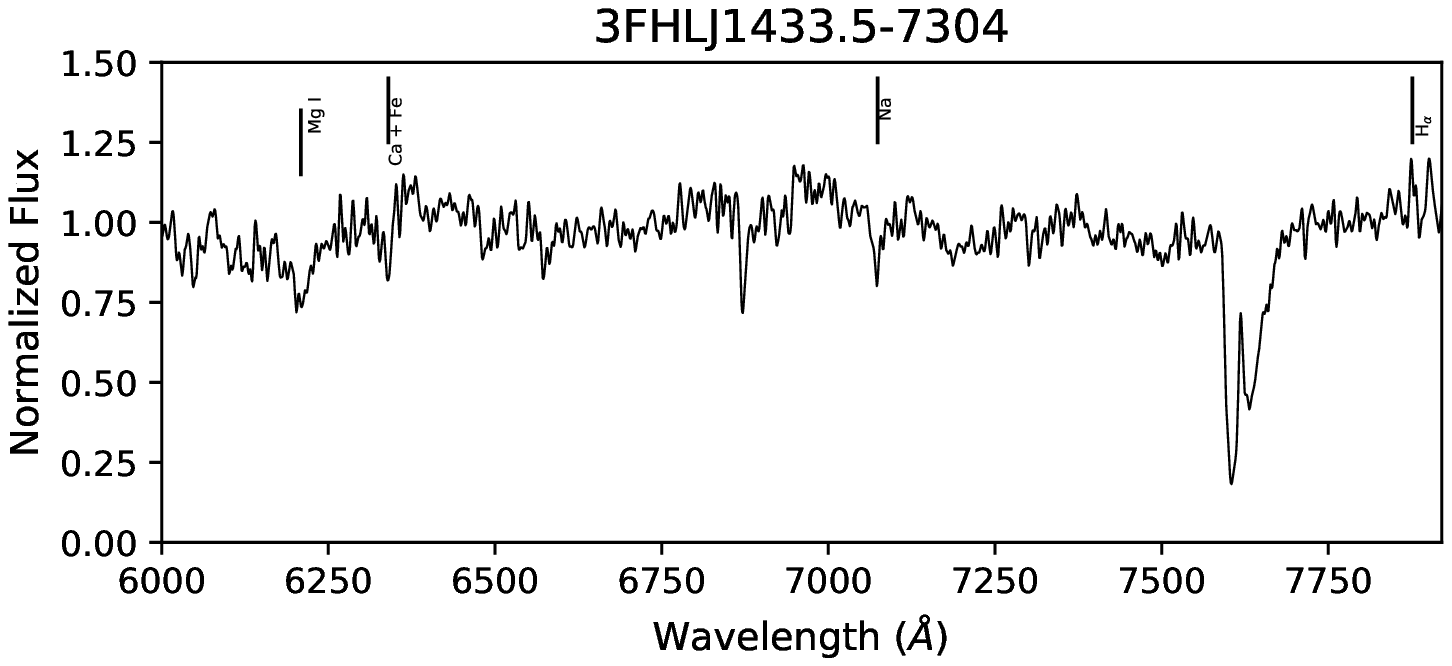}
  \end{minipage}
  \begin{minipage}[b]{.5\textwidth}
  \centering
  \includegraphics[width=0.9\textwidth]{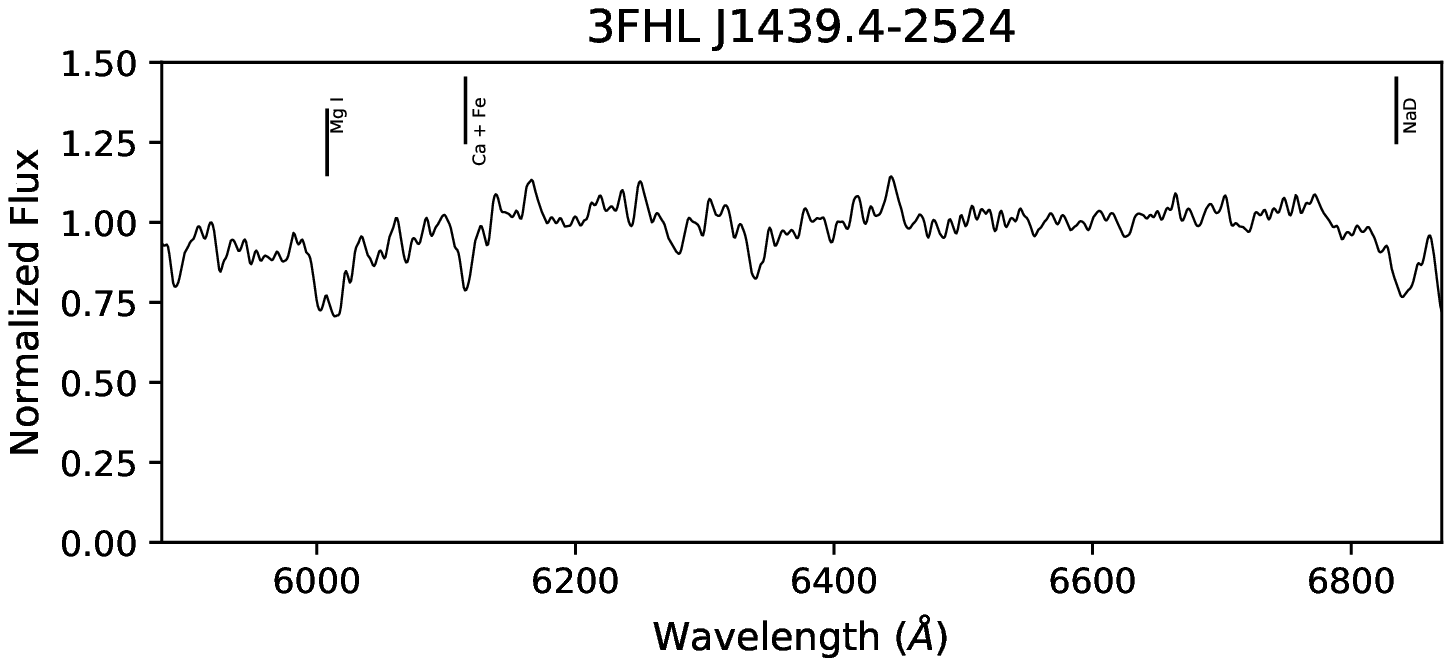}
  \end{minipage}
  \begin{minipage}[b]{.5\textwidth}
  \centering
  \includegraphics[width=0.9\textwidth]{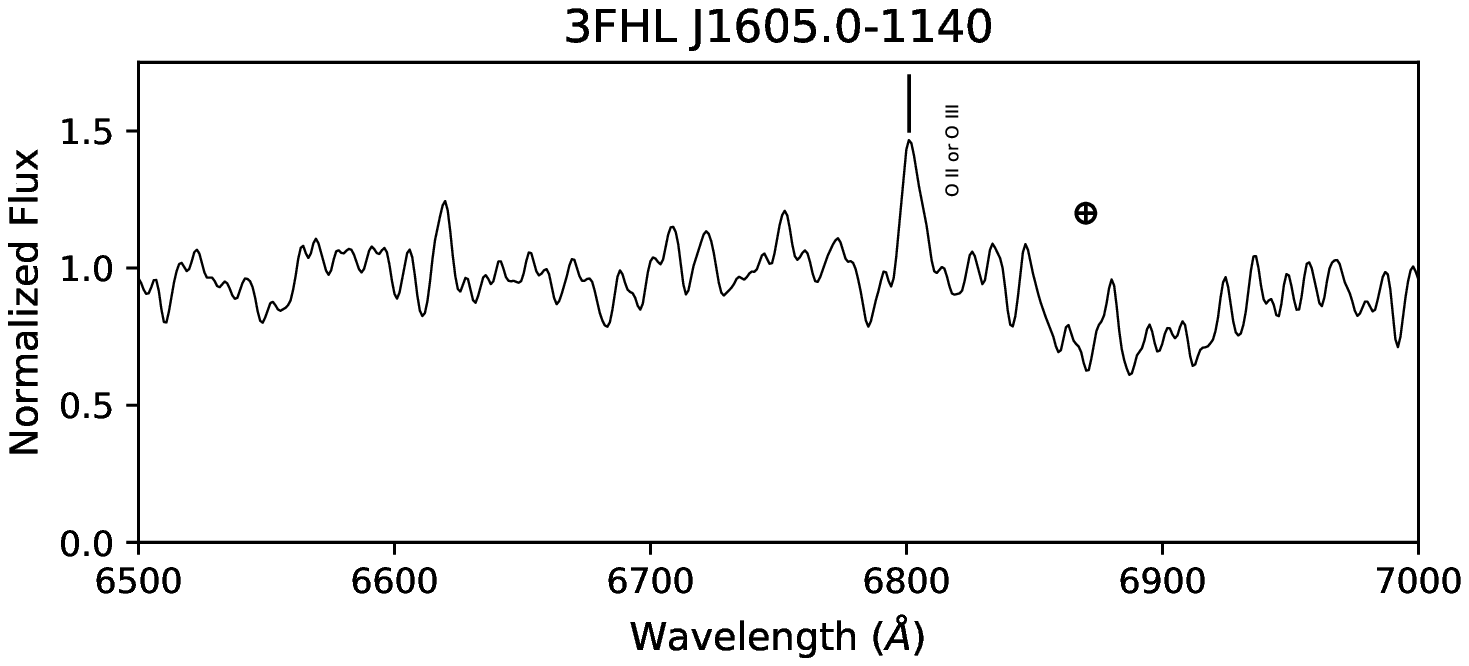}
  \end{minipage}
 
\caption{The zoomed spectra of selected sources from Fig~\ref{fig:spec} are shown above to highlight absorption and emission features }
\end{figure*}

\bibliographystyle{aasjournal}

\end{document}